\documentclass[aps,prc,floatfix,showpacs,twocolumn, preprintnumbers]{revtex4-1}

\usepackage{caption}
\usepackage{subcaption}
\usepackage{graphicx}
\usepackage{amssymb}
\usepackage{amsmath}
\usepackage{cancel}
\usepackage{placeins}
\usepackage[dvipsnames,usenames]{xcolor}
\usepackage{dcolumn}
\usepackage{bm}
\usepackage{physics}
\usepackage{hyperref}
\usepackage{float}
\usepackage{gensymb}
\usepackage{soul}

\newcommand{\carbon}{$^{12}\text{C}$}
\newcommand{\helium}{$^{4}\text{He}$}

\begin{document}

\preprint{LA-UR-24-26484}

\title{Quantum Monte Carlo calculations of electron scattering from \carbon\, in the Short-Time Approximation}
\author{L.\ Andreoli$^{1,2,3,4}$}
\email{landreol@odu.edu}
\author{G.\ B.\ King$^{3}$}
\email{kingg@wustl.edu}
\author{S.\ Pastore$^{3,4}$}
\email{saori@wustl.edu}
\author{M.\ Piarulli$^{3,4}$}
\email{m.piarulli@wustl.edu}
\author{J.\ Carlson$^{5}$}
\email{carlson@lanl.gov}
\author{S.\ Gandolfi$^{5}$}
\email{stefano@lanl.gov}
\author{R. B.\ Wiringa$^{6}$}
\email{wiringa@anl.gov}
\affiliation{
$^1$\mbox{Department of Physics, Old Dominion University, Norfolk, VA 23529, USA}\\
$^2$\mbox{Theory Center, Jefferson Lab, Newport News, VA 23610, USA}\\
$^3$\mbox{Department of Physics, Washington University in Saint Louis, Saint Louis, MO 63130, USA}\\
$^4$\mbox{McDonnell Center for the Space Sciences at Washington University in St. Louis, MO 63130, USA}\\
$^5$\mbox{Theoretical Division, Los Alamos National Laboratory, Los Alamos, NM 87545, USA}\\
$^6$\mbox{Physics Division, Argonne National Laboratory, Argonne, IL, 60439, USA}
}

\date{\today}

\begin{abstract}
The Short-Time approximation is a method introduced to evaluate electroweak nuclear response for systems  with~$A\geq12$, extending the reach of first-principle many-body Quantum Monte Carlo calculations. Using realistic two- and three-body nuclear interactions and consistent one- and two-body electromagnetic currents, we calculate longitudinal and transverse response densities and response functions of \carbon. We compare the resulting cross sections with experimental data for electron-nucleus scattering, finding good agreement.
\end{abstract}

\maketitle

\section{Introduction}

 The coming online of next-generation neutrino-oscillation experiments~\cite{DUNE:2016hlj,dune_web,DUNE:2022aul,Hyper-Kamiokande:2022smq}, including the Deep Underground Neutrino Experiment (DUNE), has brought a new wave of interest in first-principle calculations of nuclear responses. DUNE is designed to function within a broad range of neutrino energies, up to ~10 GeV, where different reaction mechanisms are at play, including the quasi-elastic, resonance production, and deep inelastic scattering regimes~\cite{Formaggio:2012,Ruso:2022qes}. A solid understanding of all these processes is required for a reliable interpretation of the experimental data.
 Specifically, accurate calculations of neutrino-nucleus cross sections are needed to reconstruct the energy of the incoming neutrinos entering the neutrino oscillation probabilities, and thus reliably extract neutrino oscillation parameters. Though computationally expensive, {\it ab initio} nuclear methods--including the Quantum Monte Carlo (QMC) methods adopted in the present work--retain the complexity of many-nucleon dynamics through the use of realistic many-nucleon interactions, as well as one- and two-body electroweak currents. The latter describe, respectively, the interaction of nucleons and pairs of correlated nucleons with external probes, such as electrons and neutrinos. 

In this work, we focus on electron-nucleus scattering in the quasi-elastic regime. In this regime, many-body dynamics, that is nucleonic correlations and electroweak currents, have been shown to be important to explain the available experimental data. For example, {\it ab initio} calculations in light nuclei correctly reproduce the observed enhancement of the electromagnetic transverse nuclear response with respect to the longitudinal one~\cite{Carlson:1997qn,Lovato:2013cua,Bacca:2014tla,Lovato:2015qka,Lovato:2016,Pastore:2019urn, Andreoli:2022,King:2024zbv}, a phenomenon that originates from the combined effect of two-nucleon correlations and the interference between one- and two-body electromagnetic currents~\cite{Lovato:2016,Pastore:2019urn,Lovato:2023khk}.

Electron scattering shares common nuclear effects with neutrino-nucleus interactions, including multi-nucleon effects, and it comes with a wealth of experimental data without the need for leptonic energy reconstruction. In fact, in these experiments, the electrons can be collimated to produce mono-energetic beams. For these reasons, electron-nucleus scattering data are extensively used to validate and test many-body nuclear models.
The synergy between electron and neutrino scattering processes has been, for example, recognized by recently established collaborations~\cite{e4nu,ntnp}, aimed at fostering collaborative efforts within the electron and neutrino physics communities.

The QMC effort has been extensively directed to calculations of inclusive electroweak response functions in the quasi-elastic regime. Most notably, the Green's Function Monte Carlo (GFMC) method has been utilized to calculate $^4$He and \carbon\, response functions induced by electrons and neutrinos, leading to an excellent agreement with the available experimental data~\cite{Lovato:2016gkq,Lovato:2017cux,Lovato:2020kba}. Due to its increasing computational cost with the number of nucleons, the GFMC method is currently limited to the study of nuclei with $A\leq 12$. Additionally, it is not suitable to describe kinematic regions characterized by large values of energy and momentum transfer, which would require the implementation of relativistic corrections at the vertex, where the correlated clusters of nucleons interact with the external probe. 

An alternative approach that addresses these shortcomings relies on the spectral function of the 
nucleus~\cite{Benhar:1994hw,Rocco:2015cil,Rocco:2018mwt}. This method is based on the factorization of the final hadronic states and has the advantage of being applicable to larger nuclear systems. Additionally, it can accommodate both relativistic kinematics and meson-production mechanisms~\cite{Rocco:2015cil,Rocco:2018mwt}. 

The Short-Time approximation (STA) was introduced in the context of QMC calculations to evaluate nuclear inclusive response functions and densities induced by electrons and neutrinos~\cite{Pastore:2019urn}. This approach is based on a factorization scheme that consistently retains two-nucleon dynamics, both correlations and consistent electroweak currents. At present, the STA algorithm has been implemented within the variational Monte Carlo (VMC) method~\cite{Carlson:2014vla} for calculations of electron scattering from the alpha particle~\cite{Pastore:2019urn}, $^3$H and $^3$He~\cite{Andreoli:2022}. STA calculations were found to be in excellent agreement with both the experimental data and the results from the GFMC method, a method that is computationally exact for these light systems.

The STA presents several advantages: Because of the adopted factorization scheme, it can be extended to i) phenomenologically describe exclusive processes, such as meson-production and nucleon knockout; and ii) incorporate relativistic kinematics at the vertex. Further, this algorithm can be applied to QMC methods, {\it e.g.}, the Auxiliary Field Diffusion Monte Carlo (AFDMC) method~\cite[and references therein]{Schmidt:1999, Carlson:2014vla}, suitable to treat nuclei of experimental interest with mass number $A > 12$. These accurate microscopic calculations of lepton-nucleus cross sections serve as improved inputs to the neutrino event generators used for the neutrino energy reconstruction procedures. As a proof of concept, the responses generated with the STA for electrons scattering from the alpha particle have been used within the GENIE event generator~\cite{Andreopoulos:2009rq} and fully tested against the world quasi-elastic electromagnetic data~\cite{Barrow:2021}, thus paving the way for analogous developments for neutrino-nucleus cross sections.

In this work, we tackle, for the first time, systems with $A=12$ -- that is, $^{12}$C -- implementing the STA within the VMC method, and present results for nuclear response densities, response functions and cross sections.
The structure of this paper is as follows: in Sec.~\ref{sec:VMC}, we provide the nuclear Hamiltonian and the electromagnetic currents used in our calculation, and describe the VMC method used to generate the ground state of $^{12}$C. In Sec.~\ref{sec:STA}, we introduce the formalism of the STA, used to evaluate response densities. The interpolation scheme developed to generate response functions on an arbitrarily fine grid of momentum transfer is presented in Sec.~\ref{sec:cross}. We summarize our results in Sec.~\ref{sec:results}, and present our conclusions in Sec.~\ref{sec:conclusions}.

\begin{figure}
\includegraphics[width=2.2in]{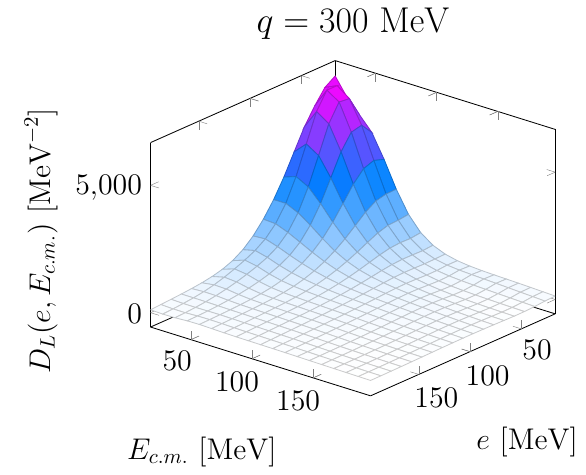}     
\caption{Total longitudinal response density of \carbon\, at {$|\mathbf{q}| = 300$ MeV/$c$}, as a function of relative energy and center of mass energy. See text for explanation.}
\label{fig:density_total}
\end{figure}
\begin{figure}
\begin{center}
\includegraphics[width=2.2in]{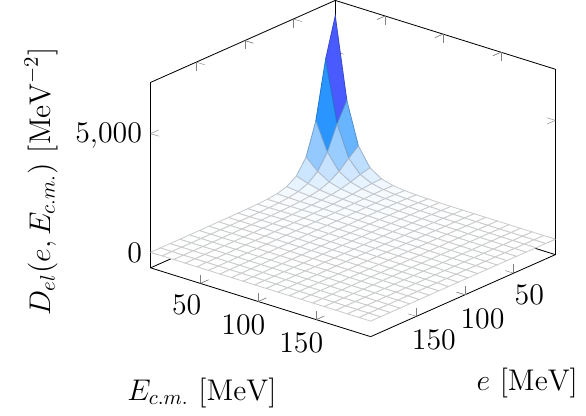}     
\end{center}
\caption{Elastic contribution to the total longitudinal response density of \carbon\, at {$|\mathbf{q}| = 300$ MeV/$c$}, as a function of relative energy and center of mass energy. See text for explanation.}
\label{fig:density_elastic}
\end{figure}

\section{Nuclear many-body Hamiltonian and electromagnetic currents}
\label{sec:VMC}

QMC methods, many-body interactions, and electroweak currents have been extensively reviewed in Refs.~\cite{Carlson:2014vla,Bacca:2014tla,Lynn:2019rdt,Gandolfi:2020pbj,King:2024zbv}. Here, we briefly sketch the VMC method, highlight the salient features of many-body interactions and currents relevant to this study, and refer the interested reader to the aforementioned review articles for details.
In this work, we use the many-body nuclear Hamiltonian
\begin{equation}
H=\sum_{i} T_{i}+\sum_{i<j} v_{i j}+\sum_{i<j<k} V_{i j k}\ ,
\label{eq:Hamiltonian}
\end{equation}
where $T_{i}$ is the single-nucleon non-relativistic kinetic energy, and $v_{i j}$ and $V_{i j k}$ denote two- and three-nucleon interaction operators, respectively. Specifically, we used the phenomenological Argonne $v_{18}$ (AV18) two-nucleon potential~\cite{Wiringa:1994wb} supplemented by the Urbana-X~\cite{Wiringa:2014} three-nucleon interaction. The AV18 consists of one-pion-exchange (OPE) contributions, and phenomenological short-to-intermediate range terms. It is fitted to the Nijmegen $np$ and $pp$ scattering database~\cite{Stoks:1993tb}, along with the deuteron binding energy, achieving a $\chi^2/$datum $\sim 1$. While the AV18 interaction is fitted to scattering data in the laboratory energy range $[0,350]$ MeV, it reproduces nucleon-nucleon phase shifts beyond the fitting range (up to $\sim 1$ GeV).

The nuclear variational wave function, $|\Psi_V\rangle$, is generated by minimizing the energy expectation value with respect to a number of variational parameters. The wave function correctly reflects the long- and short-range physics induced by the two- and three-nucleon potentials through two- and three-body correlation operators, $U_{ij}$ and $\tilde{U}^{TNI}_{ijk}$, respectively, and it reads
\begin{equation}
   |\Psi_V\rangle =
      {\cal S} \prod_{i<j}^A
      \left[1 + U_{ij} + \sum_{k\neq i,j}^{A}\tilde{U}^{TNI}_{ijk} \right]
      |\Psi_J\rangle,
\label{eq:psit}
\end{equation}
where the Jastrow wave function $\Psi_J$ is fully anti-symmetric and has the correct quantum numbers for the ground state. The overall anti-symmetry of the wave function is preserved via the symmetrization operator ${\mathcal S}$ acting on the two- and three-body correlation operators.

In order to study and predict nuclear electromagnetic cross sections, we make use of electroweak current operators that describe the interaction of nuclei with electroweak fields. In particular, the electromagnetic charge, $\rho$, and current, ${\bf j}$, are decomposed into one- and  two-body operators as
\begin{eqnarray}
\rho    &=& \sum_i {\rho}_i({\bf q}) + \sum_{i<j} {\rho}_{ij}({\bf q})   , \\
\nonumber
{\bf j} &=&  \sum_i {\bf j}_i({\bf q}) + \sum_{i<j} {\bf j}_{ij}({\bf q})    \ ,
\end{eqnarray}
where ${\bf q}$ is the momentum transferred to the nucleus and the index $i$ denotes a nucleon. The single nucleon charge and current operators ($\rho_i$, ${\bf j}_i$) are obtained from a non-relativistic reduction of the nucleon electroweak covariant currents~\cite{Carlson:1997qn}, and are written in terms of the nucleonic form factors required to correctly reproduce fall-off at increasing values of three-momentum transfer. 

The two-body operators ($\rho_{ij}$, ${\bf j}_{ij}$) used in this work have been most recently summarized in Refs.~\cite{Schiavilla:1989,Marcucci:2005zc}, and consist, primarily, of contributions of one-pion-exchange (OPE) range. More specifically, the two-body electromagnetic current operator, ${\bf j}_{ij}$, consists of model-independent and model-dependent terms, the former being constructed by requiring that they satisfy the current conservation relation within the AV18. In this sense, two-body currents are consistent with the nucleon-nucleon interaction, in that their behavior at both short and long ranges is consistent with that of the two-nucleon correlations. The model-dependent terms cannot be constrained via current conservation. The dominant contribution to the model-dependent component is associated with the excitation of intermediate (virtual) $\Delta$-isobars; in this type of contribution, the external probe excites the nucleon to a $\Delta$ that decays emitting a pion which is reabsorbed by another nucleon~\citep{Marcucci:2005zc,Schiavilla:1992sb}. The two-body charge operator, ${\rho}_{ij}({\bf q})$, consists of one-pion range terms, 
which provide contributions with a size comparable to that of a (small) relativistic correction.
Incidentally, the observed large excess of the electromagnetic transverse strength is primarily due to the interplay between one- and two-nucleon currents with two-nucleon correlations of one-pion-range~\cite{Pastore:2019urn}, a dynamical feature we need to preserve when developing approximate algorithms for larger nuclear systems.

\begin{figure}
\centering
\includegraphics[width=3.5 in]{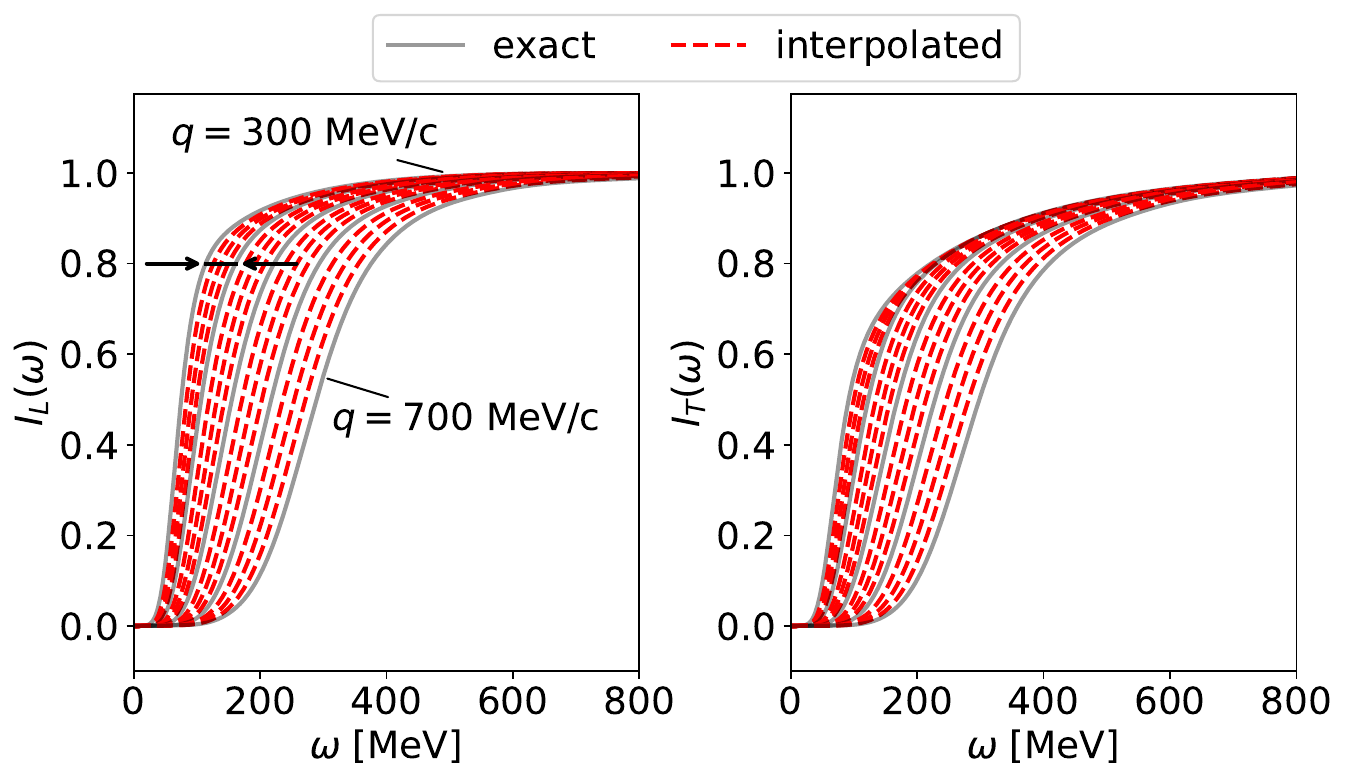}
\caption{Calculated (gray) and interpolated (red) integral functions of \helium\, at values of $|\mathbf q|$ in the range $[300, 700]$ MeV/$c$. Each integral function is obtained by interpolating neighboring exact integral functions corresponding to values of $|\mathbf q|$ spaced by $100$ MeV/$c$. }
\label{fig:interpolation_sum_rules}
\end{figure}
\begin{figure}
\centering
\includegraphics[width=2.6 in]{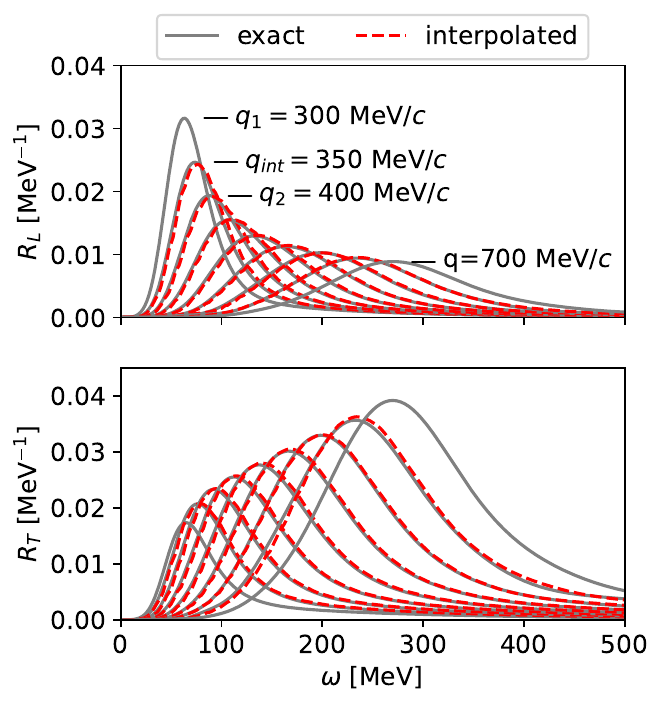}
\caption{Exact (gray) and interpolated (dashed red) longitudinal (upper panel) and transverse (lower panel) response functions for \helium, calculated for values of $|\mathbf q|$ in the range of $[300,700]$ MeV$/c$, with $50$ MeV$/c$ spacing. Each response is obtained by interpolating neighboring exact responses, corresponding to values of $|\mathbf q|$ spaced by $100$ MeV$/c$. See text for details. }
\label{fig:interpolation}
\end{figure}

\section{The Short-Time Approximation}
\label{sec:STA}

We express the cross section for inclusive electron scattering on a nucleus in terms of longitudinal and transverse nuclear response functions of the form
\begin{align}	
       R_\alpha (\mathbf{q},\omega) &= 
	{\overline{\sum_{M_i}}} \sum_f   \mel{\Psi_i}{O_\alpha ^\dagger (\mathbf{q})}{\Psi_f}	 
		 \mel{\Psi_f}{O_\alpha ({\bf q})}{\Psi_i} \, \nonumber \\
		 &\times \delta(E_f - E_i - \omega)\ ,
\label{eq:responses}
\end{align}
where $\omega$ and ${\bf{ q}}$ are the energy and momentum carried by the probe, $R_L$ is the longitudinal component ($\alpha=L$) induced by the electromagnetic charge operator, $\mathcal{O}_L=\rho$, and $R_T$ is the transverse component ($\alpha=T$) induced by the electromagnetic (vector) current operator, $\mathcal{O}_T={\bf j}$.
$\left|\Psi_i\right\rangle$ is the VMC ground state of the system with energy $E_i$ and spin $J_i$, and $\left|\Psi_f\right\rangle$ is the final state with energy $E_f$. An average of the projections $M_i$ of the initial state with spin $J_i$ is understood.

Without loss of generality, the response  can  be  equivalently  written  as  the  matrix  element  of  a  current-current correlator by replacing the sum over final states appearing in Eq.~(\ref{eq:responses})  with a real-time propagator, as
\begin{eqnarray}   
 R_\alpha (\mathbf{q},\omega) &=& 
  \int_{-\infty}^\infty  \frac{d t}{2 \pi} \,{\rm e}^{ i \left(\omega+E_i\right)   t } \nonumber \\
  && \overline{\sum_{M_i}} \,\mel{\Psi_i}{O_\alpha^\dagger (\mathbf{q})\,{\rm e}^{-i Ht}\,
 	  O_\alpha (\mathbf{q})}{\Psi_i} \ .
\label{eq:realtime}
\end{eqnarray}	
The STA uses a factorization scheme in which two-body physics is retained in both currents and strong interaction.
Hence, in the equation above, $H \approx H_{12} =  T_1+T_2 + v_{12}$ only includes up to two-nucleon interactions (in our case, the AV18). The {\it{final states}} considered here are only those with correlated nucleon pairs interacting with the external probe. This leads to a reduced computational cost compared to calculations where the full $A$-nucleon system is propagated, with pairs and triplets participating in the scattering process. For example, for \carbon\ we found a reduced factor of $\sim 10$ difference in computing time with respect to exact GFMC calculations.

\begin{figure}
\includegraphics[width=2.90 in]{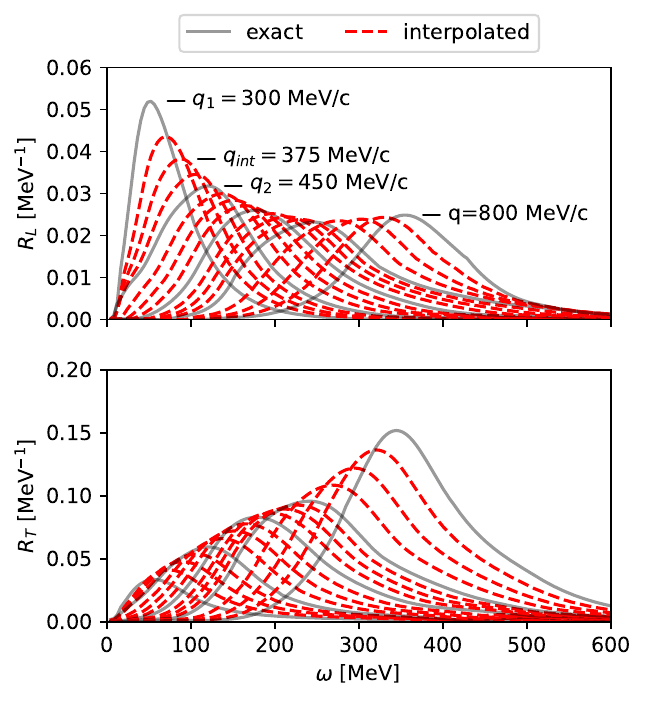} 
\caption{Exact (gray) and interpolated (dashed red) longitudinal (upper panel) and transverse (lower panel) response functions for \carbon, calculated for values of $|\mathbf q|$ in the range of $[300,800]$ MeV$/c$.}
\label{fig:c12interpolation}
\end{figure}

Correlating at most two nucleons at a time requires that we retain terms linear in $t$ when performing the short-time expansion of the propagator. Including the full two-nucleon propagator also includes ladder diagrams where pairs of nucleons scatter multiple times. Higher order terms will involve the propagation of three or more nucleons. This procedure ensures the conservation of sum rules and energy-weighted sum rules. 
At the next order, we are dropping terms quadratic in $t$ which have the form $T_i\,v_{jk}$ and $v_{ij}\,v_{kl}$, including permutations of the indices. The terms we drop are at most of the order of $\sim \mathcal{O}((\langle T_{\rm N}\rangle\,t)^2)$, where $\langle T_{\rm N}\rangle$ is the average kinetic energy per particle. In a simple, non-relativistic Fermi gas approximation, this is related to the Fermi energy $E_F$ of the system $\langle T_{\rm N} \rangle = 3E_F/5 \approx 30~{\rm MeV}$. We can use this value to roughly gauge the kinematics where the STA is valid. Note that quasi-elastic physics peaks at an energy transfer 
\begin{equation}
\label{eq:omegaqe}
\omega_{\rm qe} =  \sqrt{m^2+|{\bf{q}}|^2} - m \rightarrow \frac{|{\bf{q}}|^2}{2\,m} \, , 
\end{equation}
where $m$ is the nucleon mass, and we explicitly give the non-relativistic expression of $\omega_{\rm qe}$ for the purpose of this discussion. The associated timescale, $t_{\rm qe} = \omega_{\rm qe}^{-1}$, constrains the values of $|{\bf{q}}|$ where the STA is a valid description of these processes, which is thus roughly $E_F/\omega_{\rm qe} \ll 1$, or, equivalently, $|{\bf q}|\gg k_F$, with $k_F$ being the Fermi momentum. For sufficiently low values of $|{\bf{q}}|$, the region dominated by low-energy nuclear phenomena---such as collective excitations and low-lying discrete levels---overlaps with the quasi-elastic peak. Because low-energy nuclear phenomena are not explicitly included in the STA, one should not expect the quasi-elastic STA response to be valid when the two processes compete and potentially interfere. Physically, the lower limit in $|{\bf{q}}|$, appearing when only two nucleons at a time are correlated, arises from the absence of quantum interference effects and Pauli blocking required to describe physics close to the nuclear Fermi surface, characterized by $k_F \approx 270$ MeV/$c$. 

With the connection between short-time and propagating only two active nucleons now established, we may consider the form of the response function under such an approximation. Restricting the current-current correlator to terms containing at most two active nucleons in the final states, the correlator appearing in Eq.~(\ref{eq:realtime}) becomes:
\begin{eqnarray}
\label{eq:expansion}
&&O^\dagger\, {\rm e}^{-iHt}\,  O
=\sum_i O_i^\dagger\,  {\rm e}^{-iHt} O_i 
  + \sum_{i \neq j} O_i^\dagger\,  {\rm e}^{-iHt} O_j \\
&& + \sum_{i\neq j}
  \left(\! O^\dagger_{i}{\rm e}^{-iHt} O_{ij} + O^\dagger_{ij}{\rm e}^{-iHt} O_{i}   + O^\dagger_{ij}{\rm e}^{-iHt} O_{ij}\!\right) \, ,\nonumber
\end{eqnarray}
where the index $\alpha$ has been dropped for convenience and terms including a sum over $i\neq j\neq k$ are neglected.
From the equation above, it is apparent that, while three-nucleon effects are not accounted for in the final state (although they are fully included in the ground state), interference terms between one- and two-nucleon currents along with the AV18 two-nucleon correlations are consistently retained within the STA. In the remainder of this work, by `one-body' contribution we indicate the sum of the first two terms in Eq.~(\ref{eq:expansion}), that correspond to the one-body diagonal and one-body off-diagonal contributions, respectively. With `two-body' contributions we indicate the sum of the terms appearing in the parenthesis. These corrections are dominated by the one- and two-body interference contributions (first two terms in the parenthesis), supplemented by a small correction coming from the pure two-body operators (last term in the parenthesis). 
\begin{figure*}
\centering
    \begin{subfigure}{0.32\textwidth}
    \includegraphics[width=\hsize]{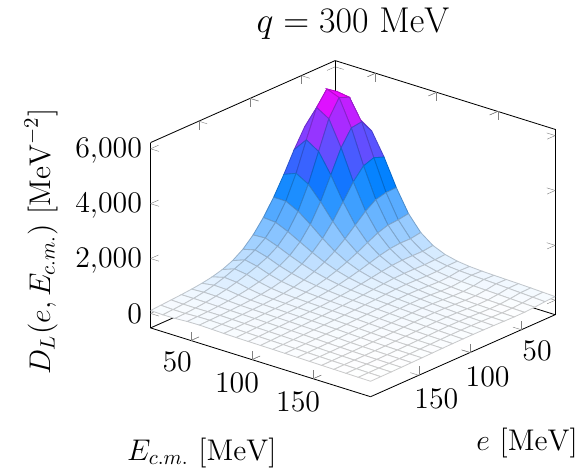}
    \subcaption{}\label{fig:densitiesa}
    \end{subfigure}
    \begin{subfigure}{0.32\textwidth}
    \includegraphics[width=\hsize]{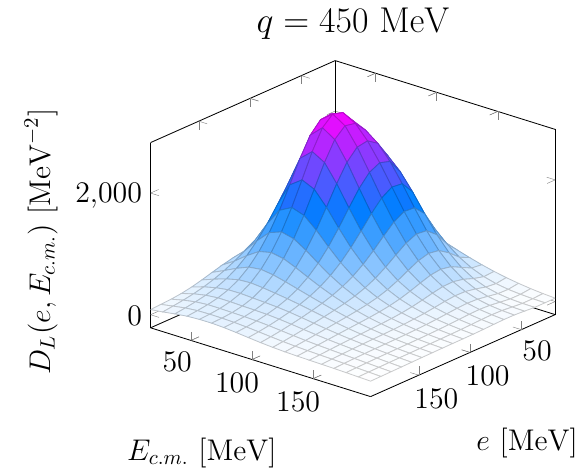}
    \subcaption{}\label{fig:densitiesa}
    \end{subfigure}
    \begin{subfigure}{0.32\textwidth}
    \includegraphics[width=\hsize]{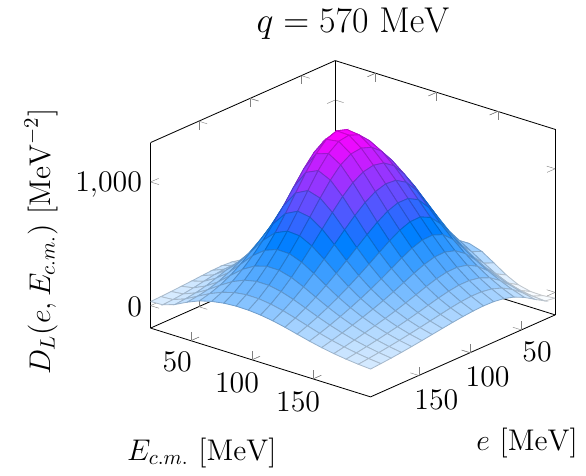}
    \subcaption{}\label{fig:densitiesa}
    \end{subfigure}
    \begin{subfigure}{0.32\textwidth}
    \includegraphics[width=\hsize]{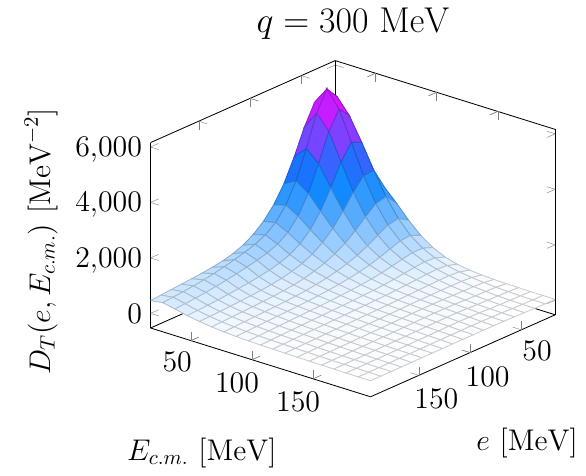}
    \subcaption{}\label{fig:densitiesa}
    \end{subfigure}
    \begin{subfigure}{0.32\textwidth}
    \includegraphics[width=\hsize]{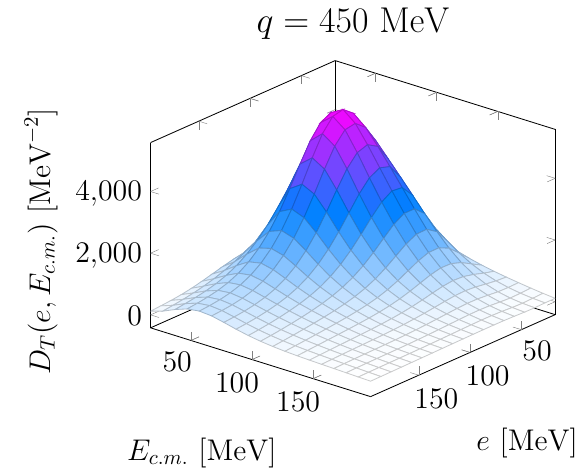}
    \subcaption{}\label{fig:densitiesa}
    \end{subfigure}
    \begin{subfigure}{0.32\textwidth}
    \includegraphics[width=\hsize]{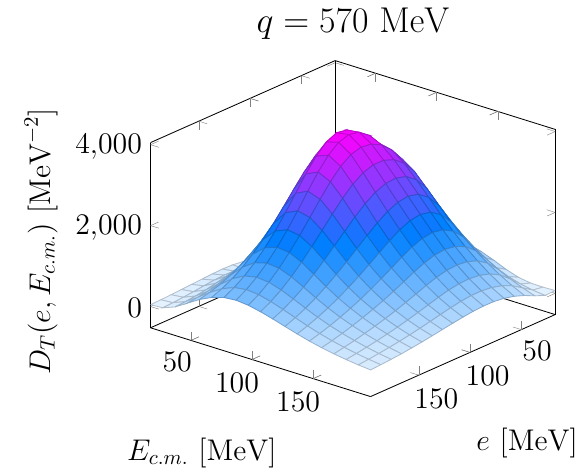}
    \subcaption{}\label{fig:densitiesa}
    \end{subfigure}
\caption{Longitudinal (top) and transverse (bottom) response densities of \carbon, for different values of $\mathbf{q}$, as functions of the center of mass energy $E_{c.m.}$ and relative energy $e$ of the struck nucleon-nucleon pair.}
\label{fig:densities}
\end{figure*}
After the insertion of a complete set of two nucleon states~\cite{Pastore:2019urn}, 
the response in~Eq.(\ref{eq:realtime}) can be evaluated as an integral of a 
response density, $D_\alpha(e,E_{\rm c.m.})$, over the relative energy, $e$, and center of mass energy, 
$E_{\rm c.m.}$, of the interacting pair (or, equivalently, over relative and CM momenta, ${\bf p}^\prime$ and ${\bf P}^\prime_{\rm c.m.}$, of the pair of struck nucleons). The response density gives access to explicit information about two-nucleon dynamics in the final states, and, upon its integration over the relative and center of mass energies of the active pair of nucleons, one obtains the corresponding response function, $R_\alpha ({\bf q},\omega)$, as
\begin{eqnarray}
 R_\alpha ({\bf q},\omega) &=&
  \int_{0}^{\infty} d e \int_{0}^{\infty} d E_{\mathrm{c.m.}} \,
  D_\alpha\left(e, E_{\mathrm{c.m.}}\right)\, \nonumber \\
  &\times&\,\delta\left(\omega+E_i-e-E_{\mathrm{c.m.}}\right) \ .
\label{eq:density}
\end{eqnarray}

The response density given above has purely elastic contributions coming from the ground state for which 
the associated response is $\propto |\langle \Psi_0 | O_\alpha({\bf q})| \Psi_0 \rangle|^2$. 
The elastic contribution becomes more prominent as the momentum transfer decreases, and it coincides with the fully elastic response in the limit of $|\mathbf{q}|\rightarrow 0$. In this work, we subtract the elastic contribution adopting the procedure introduced in Ref.~\cite{Andreoli:2022}. Specifically, we calculate the elastic response density $D_{\alpha}^{\rm el}$ as 
\begin{align}
&D_\alpha^{\mathrm{el}}({\bf q}, {\bf p}^\prime, {\bf P}^\prime) = \,
 \left|\left\langle \Psi_0 | O_\alpha \left(\mathbf{q}\right) \mid \Psi_0 \right\rangle\right|^{2}\nonumber \\
&\times \sum_{\beta} \langle \Psi_0| \Psi_{2}\left({\bf p}^\prime, {\bf P}^\prime_{\rm c.m.}, \beta \right)\rangle \langle \Psi_{2}\left({\bf p}^\prime, {\bf P}^\prime_{\rm c.m.}, \beta \right) |\Psi_0 \rangle \ ,
\end{align}
where  $\Psi_2$ schematically denotes intermediate states with two active nucleons, and the sum runs over all two body quantum numbers $\beta$. The expression above is obtained under the assumption that, at low values of momentum transfer, the internal nuclear dynamics of the ground state dominates~\cite{Andreoli:2022}.
At $|{\bf q}|=300$ MeV$/c$, where the elastic contribution is relevant, we subtract the longitudinal contribution $D_L^{\mathrm{el}}(e, E_{\rm c.m.})$ from the total longitudinal response density $D_L(e, E_{\rm c.m.})$.

As discussed above,  responses calculated for $|{\bf{q}}| \lesssim 300$ MeV/$c$ do not reproduce the correct threshold behavior. This is enforced by redistributing the strength of the response below a threshold $\omega_{\text{th}}$ to higher values of $\omega$.
This is achieved through the insertion of a Gaussian distribution controlled by a width parameter $\overline{\omega}$, and imposing that the sum rule is preserved. A detailed description of this procedure can be found in Ref.~\cite{Pastore:2019urn}.
In the present calculation, we use $\omega_{\text{th}}=25$ MeV and $\overline{\omega}=15$ MeV.

\section{Cross Sections and Interpolation Scheme}
\label{sec:cross}

The inclusive double differential cross section for electron-nucleus scattering is written in terms of the nuclear response functions as
\begin{equation}
\frac{d^{2} \sigma}{d E_{e^{\prime}} d \Omega_{e}} =\left(\frac{d \sigma}{d \Omega_{e}}\right)_{\mathrm{M}}\left[A_{L} R_{L}(\mathbf{q}, \omega)+A_{T} R_{T}(\mathbf{q}, \omega)\right]\ ,
\label{eq:cross_section}
\end{equation}
where the leptonic kinematic factors read
\begin{equation}
A_{L}=\left(\frac{q^{2}}{\mathbf{q}^{2}}\right)^{2}\ , \quad A_{T}=-\frac{1}{2} \frac{q^{2}}{\mathbf{q}^{2}}+\tan ^{2} \frac{\theta_{e}}{2} \, ,
\label{eq:al_at}
\end{equation}
while the Mott cross section is defined as 
\begin{equation}
\left(\frac{d \sigma}{d \Omega_{e}}\right)_{\mathrm{M}}=\left[\frac{\alpha \cos \left(\theta_{e} / 2\right)}{2 E_{e^{\prime}} \sin ^{2}\left(\theta_{e} / 2\right)}\right]^{2} \, . 
\end{equation}
In the equations above, $\alpha$ is the fine structure constant and $q^2=\omega^2-|\mathbf q|^2$, while $E_{e'}$ is the final electron energy, and $\theta_e$ is the electron scattering angle.

\begin{figure}
\centering
\includegraphics[width=2.4in]{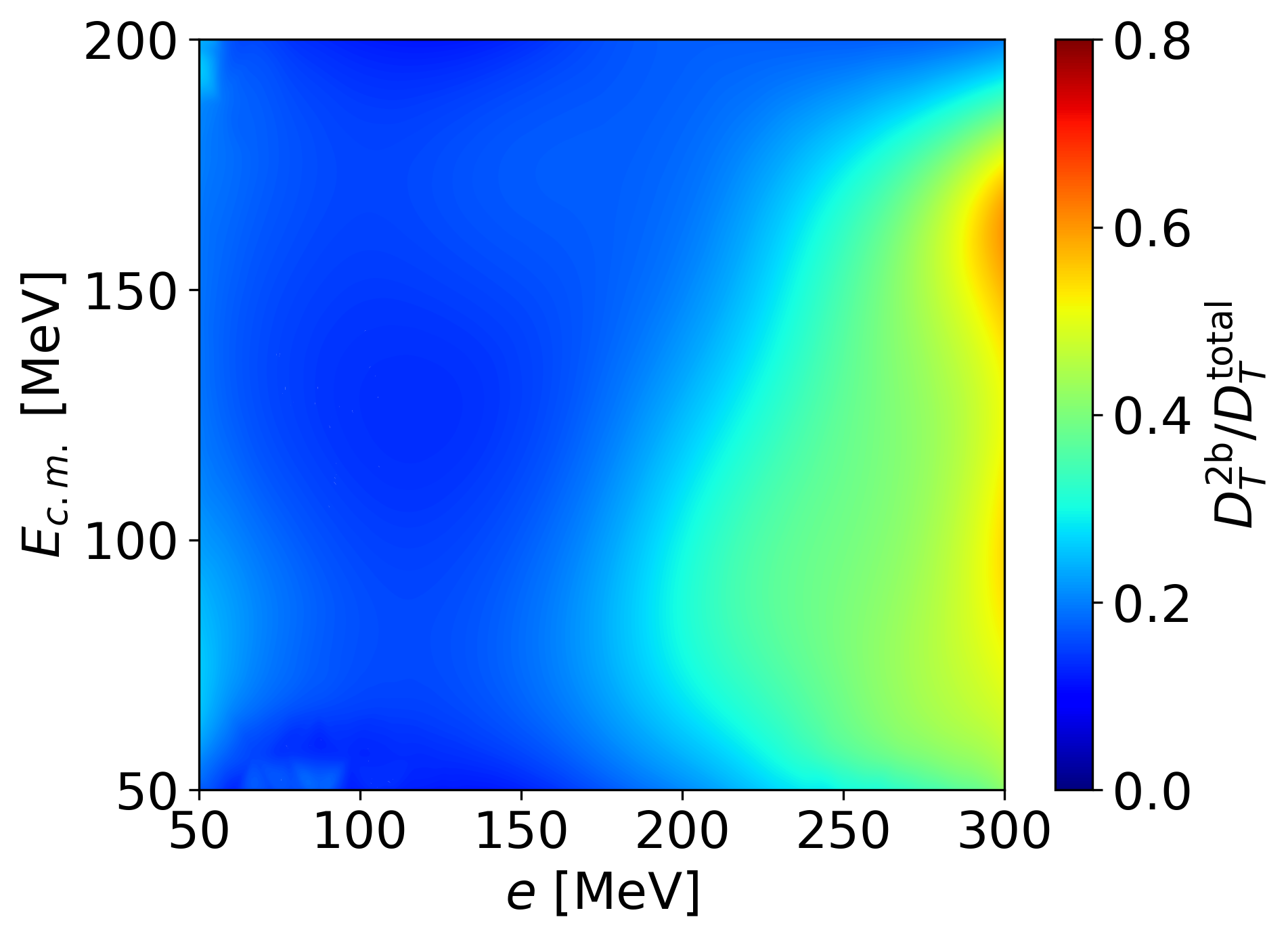}  
\caption{Two-body contributions (interference plus pure two-body) to the total transverse response density, $D_{T}^{\rm 2b}/D_{T}^{\rm total}$, at $|\mathbf{q}|=570$ MeV/$c$. See text for details.}
\label{fig:2body}
\end{figure}

While the STA algorithm allows for a reduced computational cost,  calculations of \carbon's nuclear response densities within the STA are still demanding,
which prevents from obtaining response functions calculated over a finely-spaced grid in momentum transfer. These are required to accurately compute the cross sections. Several interpolation schemes~\cite{Rocco:2018tes,Barrow:2021} have been adopted to generate response functions from the calculated and sparse ones. For example, the approach of Ref.~\cite{Rocco:2018tes} exploits the scaling behavior of the response functions to generate response functions for arbitrary values of $\omega$ and $\mathbf{q}$. 

In this work, we develop an interpolation scheme that conserves the sum rules. Specifically, we calculate the normalized cumulative integral of the response function, $I_{L/T}$, at fixed value of momentum transfer and as a function of omega, 
\begin{equation}
\label{eq:integral}
I_{L / T}(\omega; \mathbf{q}_i)=\frac{\int_0^\omega R_{L / T}\left(\omega^{\prime}; \mathbf{q}_i \right) d \omega^{\prime}}{\int_0^\infty R_{L / T}\left(\omega^{\prime} ; \mathbf{q}_i \right) d \omega^{\prime}}\, ,
\end{equation}
and obtain a set of $I_\alpha(\omega; \mathbf{q}_i)$ for each of the response functions calculated within the STA at values of momentum transfer ${\bf q}_i$. These are displayed in Fig.~\ref{fig:interpolation_sum_rules} for $^4$He  (solid lines in the figure) for values of momentum transfer in the range of $|\mathbf{q}_i|=[300,700]$ MeV/$c$ with a $100$ MeV/$c$ spacing. We then interpolate the $I_\alpha(\omega; \mathbf{q}_i)$'s to obtain $I^{\rm int}_\alpha(\omega; \mathbf{q})$ for any arbitrary value of $\mathbf{q}$. Specifically, the interpolation is carried out using two integral functions at the same value, say $I_\alpha(\omega; 300\, {\rm MeV}/c)=I_\alpha(\omega; 400\, {\rm MeV}/c)=0.8$ as shown as an example in the figure, and the interpolated function, $I_\alpha^{\rm int}(\omega; \mathbf{q})=0.8$,  is evaluated for any ${\bf q}$ in the range $[300,400] \,{\rm MeV}/c$. The interpolated integral functions are shown in Fig.~\ref{fig:interpolation_sum_rules} with dashed red lines (note that, for \helium, we generated integral functions and responses over a momentum grid with $1$ MeV/$c$ spacing; in the figure, we only display three interpolated functions). The corresponding interpolated response function is recovered from $I^{\rm int}_\alpha(\omega; \mathbf{q})$ by taking its derivative. 

\begin{figure}
\centering
\includegraphics[width=0.80\columnwidth]{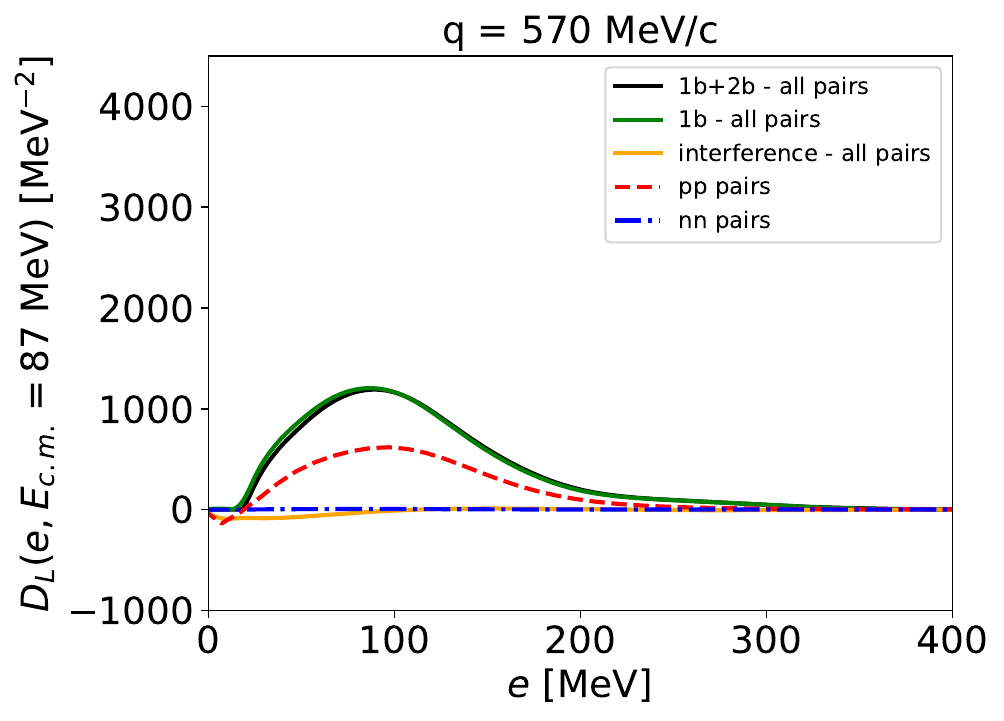} 
\includegraphics[width=0.80\columnwidth]{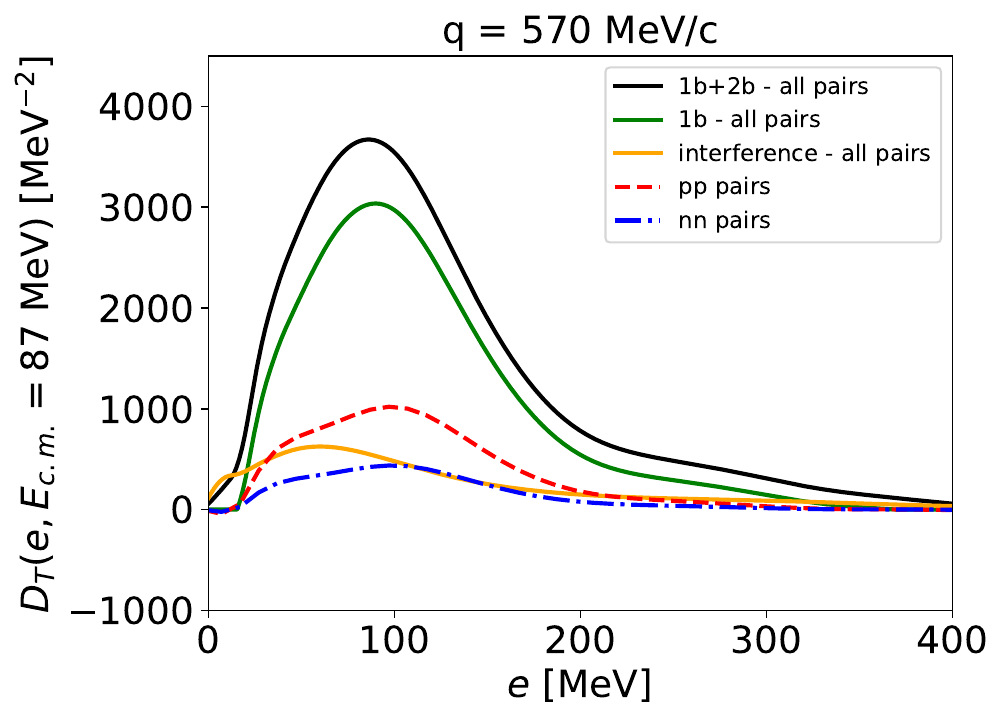}
\caption{Longitudinal (top) and transverse (bottom) response densities at $|\mathbf{q}|=570$ MeV$/c$ and $E_{\rm c.m.}=87$ MeV. Contributions from $pp$ and $nn$ pairs are shown in red dashed lines and blue dashed-dotted lines, respectively, and total responses in solid black lines. Total one-body and interference contributions are shown in solid green and orange lines, respectively.}
\label{fig:pairs}
\end{figure}

For $^4$He, the computational cost is such that calculations of nuclear responses on a fine grid of momentum transfer are feasible and can be used to validate the interpolation scheme described above. To this end, we use the STA to calculate a set of longitudinal and transverse response functions for momentum transfer in the range $|\mathbf{q}|=[300,700]$ MeV/$c$, with a $10$ MeV/$c$ spacing. We use the interpolation scheme on a limited subset of $\sim 10$ responses and check that those resulting from the interpolation procedure are in agreement with the exact ones. To further illustrate the procedure, Fig.~\ref{fig:interpolation} shows the comparison between interpolated (dashed red lines) and exact (solid gray lines) responses. In particular, the exact responses for two momenta, say $|{\bf q}_1|= 300$ MeV/$c$ and $|{\bf q}_2|= 400$ MeV/$c$ as highlighted in the figure, are used to calculate the interpolated response in the midpoint $|{\bf q}_{\rm int}|=(|{\bf q}_2| - |{\bf q}_1|)/2= 350$ MeV/$c$, which is in perfect agreement with the corresponding exact response.  

The approach described above allows us to generate longitudinal and transverse response functions using a limited number of calculated ones, while ensuring the uni-modality of the interpolated responses. Having tested the interpolation scheme on \helium, we apply it to \carbon.
Due to the aforementioned computational cost, in the case of \carbon\, we calculate the response densities and functions for only five values of momentum transfer, namely $|{\bf{q}}|=300,\, 450,\,570,\, 650$ and $800$ MeV/$c$. The associated responses are displayed in Fig.~\ref{fig:c12interpolation}
(solid gray lines) along with those obtained from them through the interpolation scheme (dashed red lines).

\begin{figure}
\includegraphics[width=2.9in]{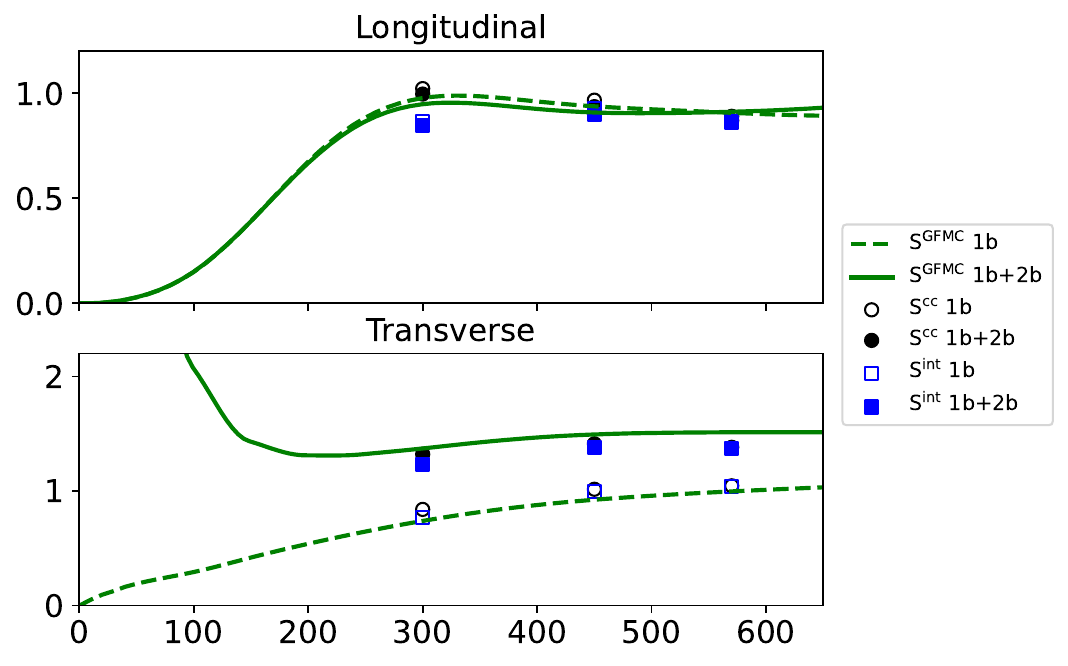}
\caption{\carbon\, longitudinal and transverse STA sum rules, S$_{L/T}^{\rm int}$ (blue squares) and S$_{\rm L/T}^{\rm cc}$ (black circles), compared with the GFMC results S$_{L/T}^{\rm GFMC}$ from Ref.~\cite{Lovato:2013cuaf} (green lines).
See text for explanations.}
\label{fig:sumrules}
\end{figure}

We conclude this section by providing the values of the longitudinal and transverse sum rules, $S_{L/T}({\bf {q}})$ in Table~\ref{tab:sumrules}. These are defined as~\cite{Pastore:2019urn}:
\begin{eqnarray}
\label{eq:sumrules}
G^2_\alpha(Q^2_{\rm qe})\, S_\alpha({\bf q}) 
& = & \int^\infty_{\omega_{\rm el}} {d \omega}\, R_\alpha({\bf q},\omega) \\
& = & \overline{\sum_{M_i}} \langle \Psi_i | 
 	  O_\alpha^\dagger ({\bf q})\, O_\alpha ({\bf q}) |\Psi_i \rangle  \ .
\end{eqnarray}
where the factor $G^2_\alpha(Q^2_{\rm qe})$ denotes the square of the appropriate combination of nucleon electromagnetic form factors~\cite{Carlson:1997qn,Lovato:2014} evaluated at
$Q_{\rm el}^2\,$=$\,{\bf {q}}^2-\omega_{\rm el}^2$. In Table~\ref{tab:sumrules}, we report the values of the sum rules obtained by i) integrating the STA response functions, as shown in the first expression in Eq.~(\ref{eq:sumrules}); and ii) calculating the current-current matrix element as illustrated in the second expression in Eq.~(\ref{eq:sumrules}) within the STA, that is neglecting three- and four-nucleon terms.
In the table, the former are denoted with `$S_\alpha^{\rm int}$' (blue squares in Fig.~\ref{fig:sumrules}) and the latter with `$S_\alpha^{\rm cc}$' (black circles in Fig.~\ref{fig:sumrules}). Values in parentheses are obtained from one-body currents alone. We compare our results with those obtained from GFMC evaluations, denoted in the table with `$S_\alpha^{\rm GFMC}$' (green lines in Fig.~\ref{fig:sumrules}).
Both sum rules, namely the $S_\alpha^{\rm int}$ (squares) and $S_\alpha^{\rm cc}$ (circles), are in agreement with the GFMC sum rules at a few percent level, with $S_\alpha^{\rm cc}$ providing a closer match to the GFMC results. The $S_\alpha^{\rm int}$ sum rules underestimate the GFMC results. In fact, the response densities are calculated up to finite values of center-of-mass and relative energies, therefore the integrated sum rules miss strength coming from contributions found beyond these finite energy ranges.
In Fig.~\ref{fig:sumrules}, with empty markers we indicate sum rules obtained with one-body operators, while those inclusive of two-nucleon currents are represented by full markers. Similarly, GFMC results based on the one-body operator are given by the dashed line, and those based on one- and two-body operators are indicated by solid lines. 

\begin{table*}
\label{tab:sumrules}
    \centering
\begin{tabular}{c | c c c | c c c}
\hline
$q$ [MeV/c] & S$_L^{\rm int}$ & S$_L^{\rm cc}$ & $S_L^{\rm GFMC}$ & $S_T^{\rm int}$  & S$_T^{\rm cc}$ & $S_T^{\rm GFMC}$ \\
\hline
300 & (0.87)0.84 & (1.02)1.00 & (0.98)0.94 & (0.77)1.23 & (0.84)1.32 & (0.74)1.37 \\
450 & (0.93)0.90 & (0.97)0.94 & (0.94)0.91 & (1.00)1.38 & (1.02)1.41 & (0.93)1.50 \\
570 & (0.88)0.86 & (0.89)0.87 & (0.91)0.91 & (1.04)1.37 & (1.05)1.38 & (1.00)1.51 \\
\hline
    \end{tabular}
    \caption{\carbon\, longitudinal and transverse STA sum rules, S$_{L/T}^{\rm int}$ and S$_{\rm L/T}^{\rm cc}$, compared with the GFMC results S$_{L/T}^{\rm GFMC}$ from Ref.~\cite{Lovato:2013cuaf}. Results based on one-body currents are given in parentheses. See text for explanations.}
    \label{tab:sumrules}
\end{table*}

\section{Results}
\label{sec:results}
In this section, we summarize our calculations. Specifically, we report the results obtained within the STA for i) response densities; ii) response functions; and iii) inclusive double differential cross sections of electron scattering from \carbon. Where available, we compare with GFMC theoretical calculations from Ref.~\cite{Lovato:2016gkq}, and experimental data from the studies of Refs.~\cite{Barreau:1983ht,JOURDAN1996117}.

\subsection{ Response Densities}
\label{sec:c12densities}

Response densities for \carbon\, are evaluated for five values of momentum transfer ($|{\bf{q}}|=300,\, 450,\,570,\, 650$ and $800$ MeV/$c$).  In Fig.~\ref{fig:densities}, we show the results obtained for $|{\bf{q}}|=300,\, 450$, and $570$ MeV/$c$. The surface plots are functions of the nucleon pair's relative, $e$, and center of mass, $E_{\rm cm}$, energies, evaluated on equally spaced grids with $10$ MeV spacing. Specifically, we use values up to $e \sim  E_{\rm cm} \sim 500$ MeV, which are sufficient to capture the main peak of the responses in the transfer momentum range we are considering here. This choice is dictated by the computational cost of the calculations and leads to missing strengths at the tails as the value of momentum transfer increases and the density's peak moves towards the edges of the region covered by the relative and center of mass energies.
At the moment, however, extending the energy range of the pair beyond $\sim 500$ MeV is computationally prohibitive.
To remedy this problem, the two-body tail of the response densities was extrapolated beyond $e=500$ MeV. The same procedure, when applied to truncated response densities for \helium\, calculated up to $e=500$ MeV, successfully reproduced the response densities calculated over the energy range $e,E=[0,900]$ MeV.

\begin{figure}
\centering
\includegraphics[width=3in]{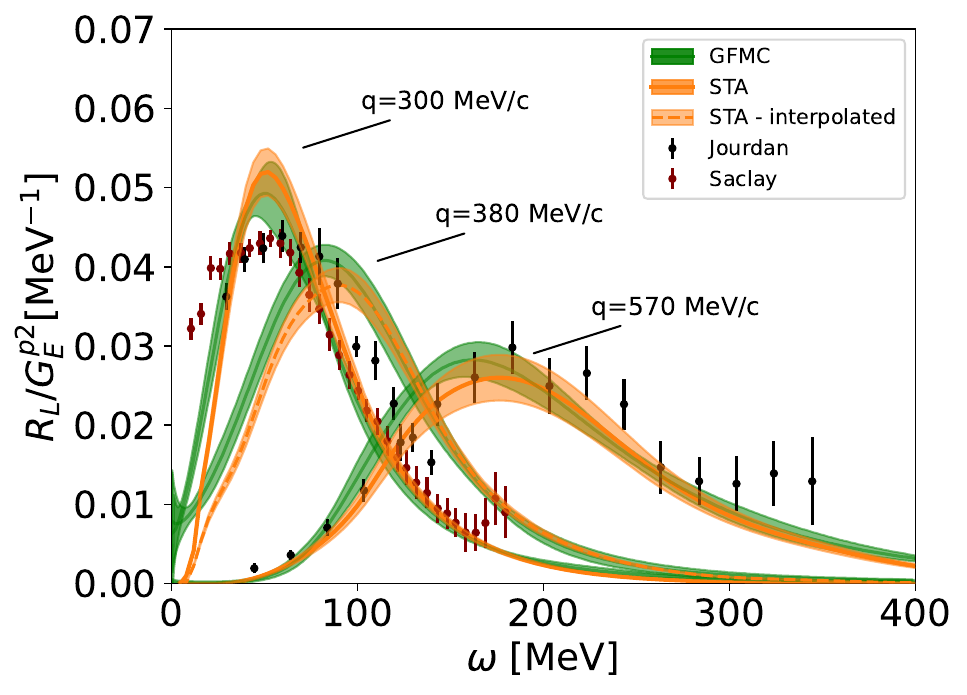} 
\includegraphics[width=3in]{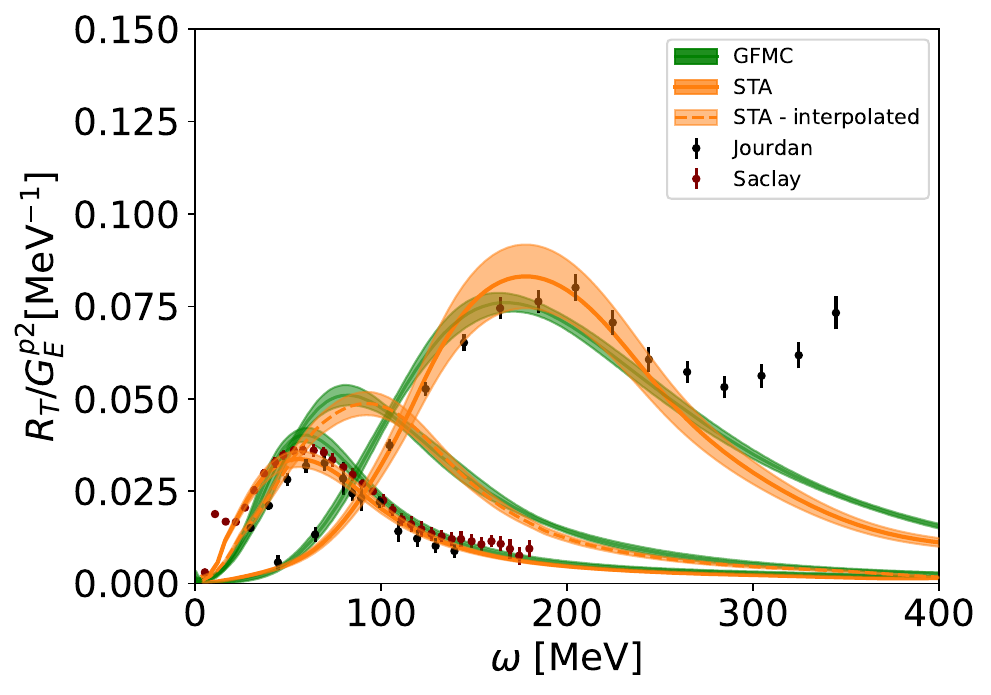} 
\caption{Longitudinal (top panel) and transverse (bottom panel) response functions for \carbon\ for different values of $|\mathbf{q}|$, obtained within the STA (solid and dashed orange), GFMC (green) from Ref.~\cite{Lovato:2016gkq}, and experimental responses obtained from~\cite{Barreau:1983ht,JOURDAN1996117}.}
\label{fig:functions}
\end{figure}

From Fig.~\ref{fig:densities}, we see that the position of the peak is realized for values of $e \sim E_{\rm cm}\sim |\mathbf{q}|^2/4\,m$, a condition resulting from simple kinematic considerations. All the densities are purposely plotted in the range $[0,200]$ MeV, for both the relative and center of mass energies, to highlight the  tail induced by two-body physics for increased values of relative energies, a finding that is in line with previous studies on two-body momentum distributions\cite{Wiringa:2014,Piarulli:2022ulk}, and STA calculations in $A=3$ and $4$ systems~\cite{Pastore:2019urn,Andreoli:2022}. Note that, the elastic contribution has been removed from the longitudinal responses displayed in the figure.

\begin{figure}
\includegraphics[width=2.9in]{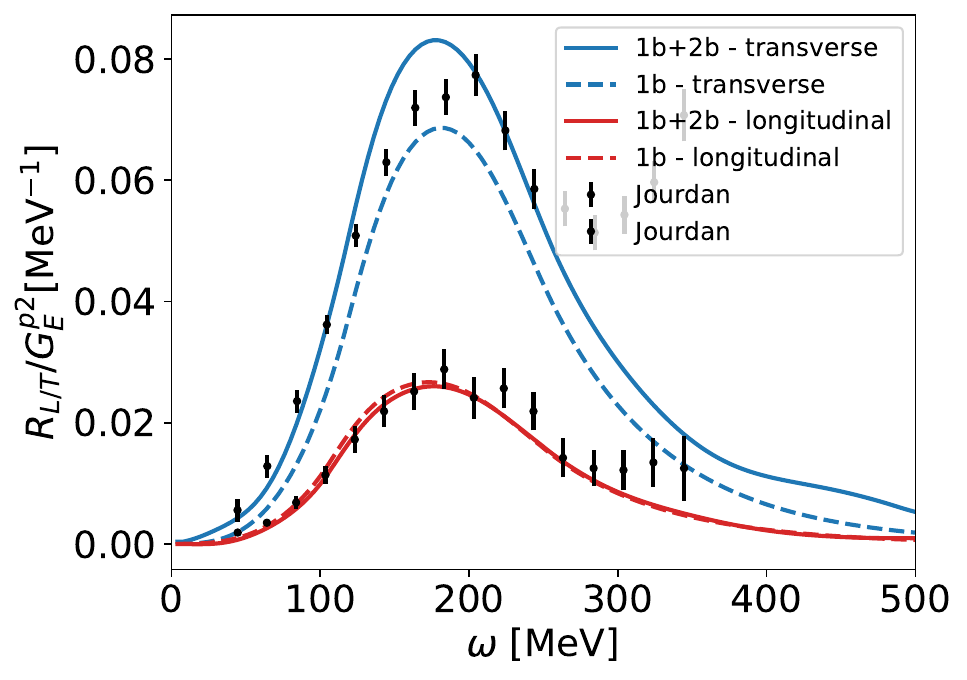}
\caption{Longitudinal (red lines) and transverse (blue lines) response functions at $|\mathbf{q}|=570$ MeV. Response functions accounting for one-body (one- plus two-body) currents are indicated with dashed (solid) lines.}
\label{fig:res_fun_2b}
\end{figure}

To further analyze the impact of two-body physics on the calculated response densities, in Fig.~\ref{fig:2body}, we show a contour plot of the two-nucleon currents' contribution (interference plus pure two-body terms) relative to the total transverse response density, {\it i.e.}, $D_{T}^{\rm 2b}/D_{T}^{\rm total}$, as a function of the relative and center of mass energies at $|\mathbf{q}|=570$ MeV. The two-nucleon currents' effect is moderate at low values of relative energy,  while it increases at higher values, providing, on average, a thirty percent correction.
For the longitudinal response, an analogous contour plot indicates that the two-nucleon currents' contribution is less than $5\%$ of the total response, as expected given the negligible relativistic correction induced by two-nucleon charge operators.

\begin{figure*}
\centering
\begin{subfigure}[t]{6in}
\includegraphics[width=\linewidth]{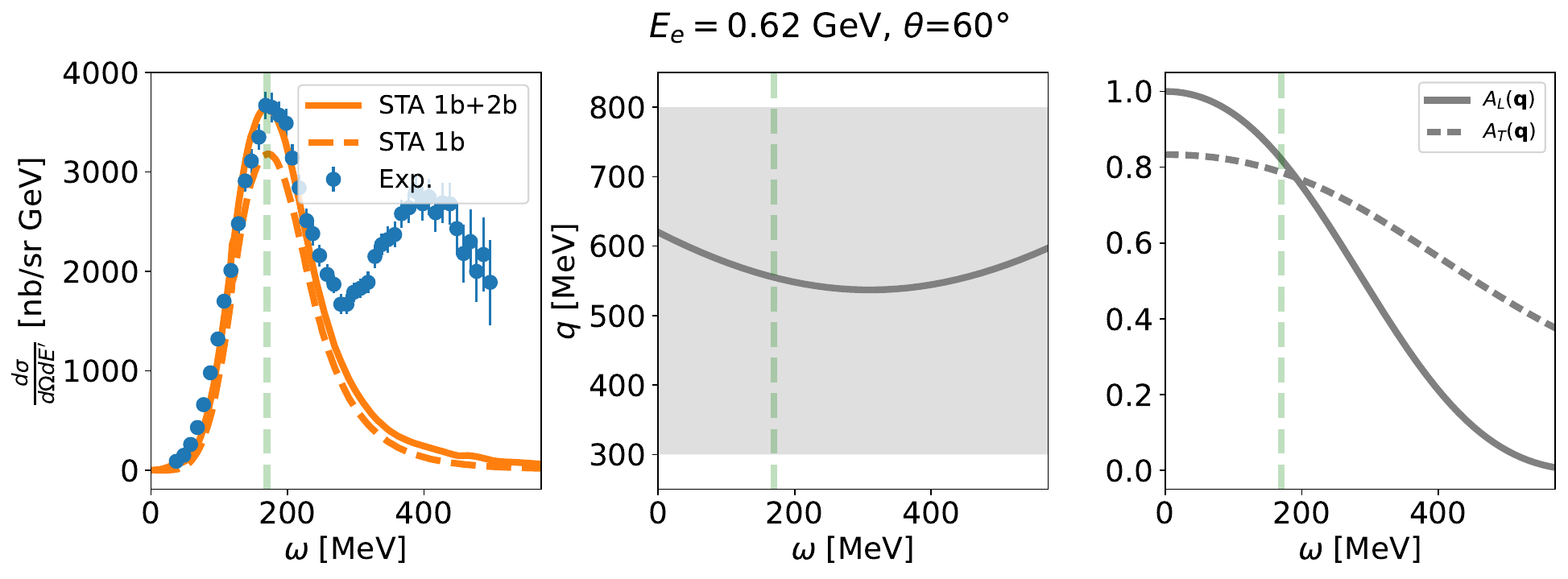}
\subcaption{}\label{fig:q_values1a}
\end{subfigure}
\begin{subfigure}[t]{6in}
\includegraphics[width=\linewidth]{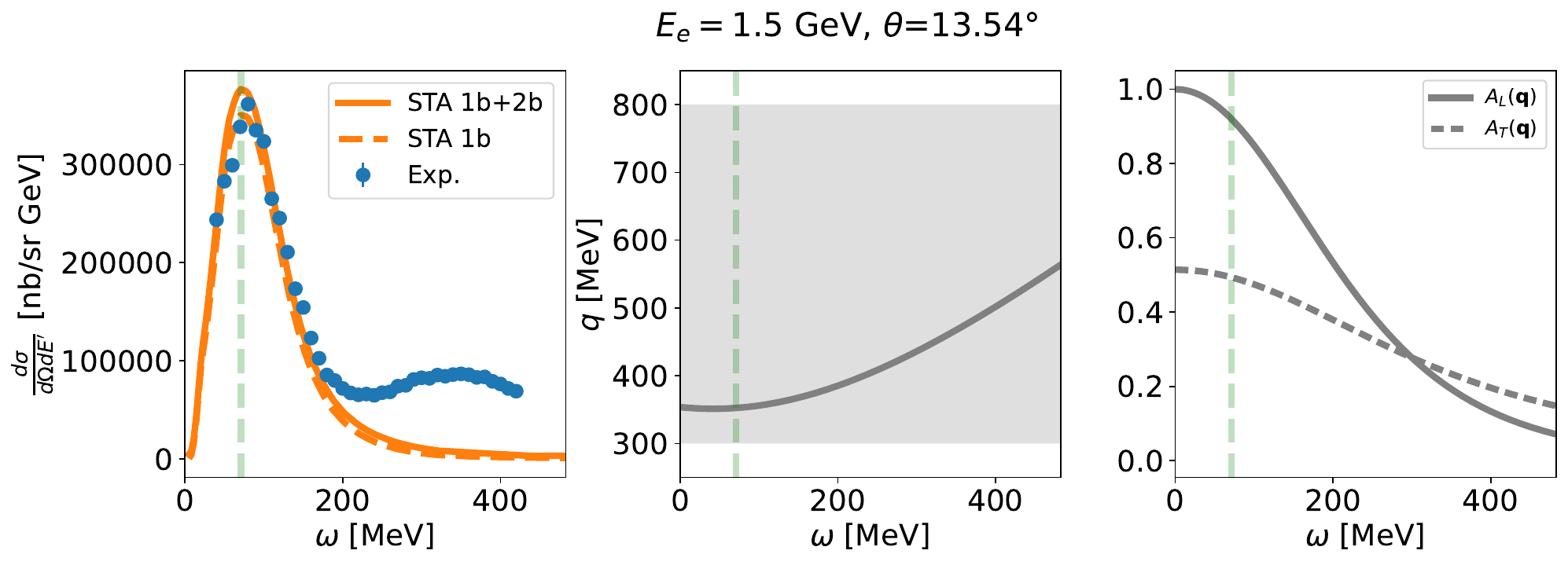}
\subcaption{}\label{fig:q_values1b}
\end{subfigure}
\begin{subfigure}[t]{6in}
\includegraphics[width=\linewidth]{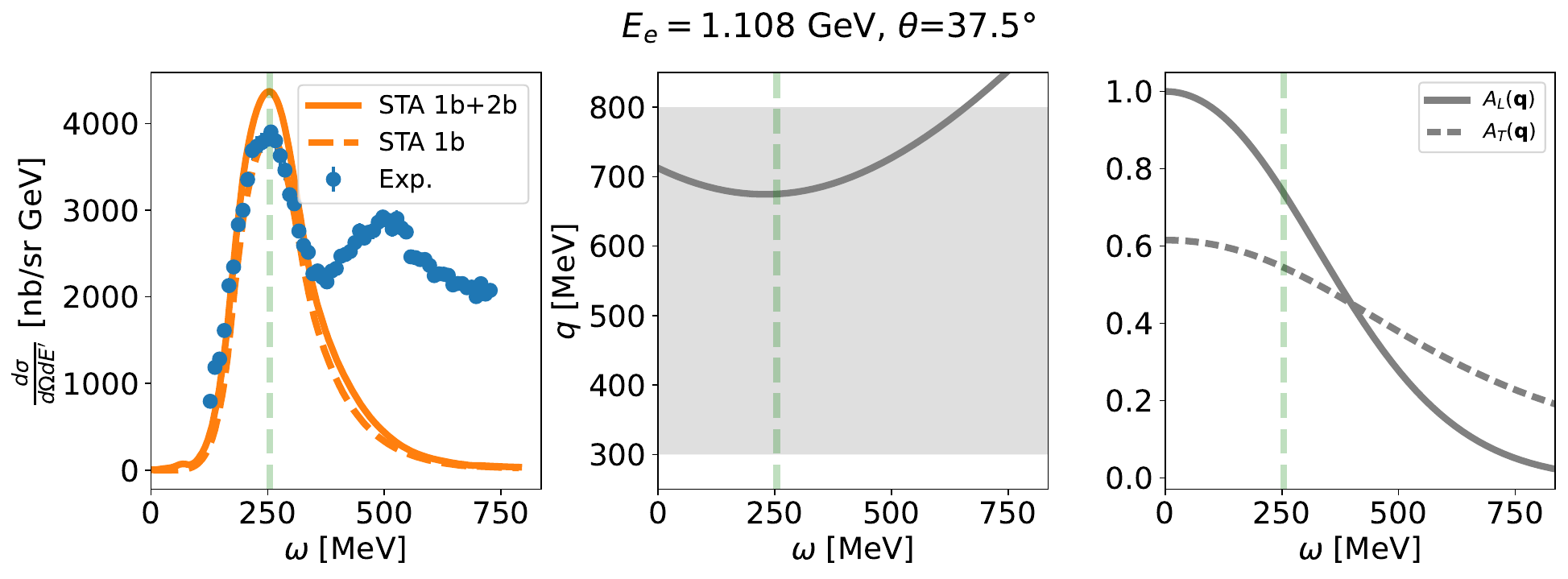}
\subcaption{}\label{fig:q_values1c}
\end{subfigure}
\caption{Inclusive double-differential cross sections for electron scattering on \carbon\, are shown in the left panels.
The values of $|\mathbf{q}|$ and $\omega$ leading to the selected electron's energy and scattering angle, $E_e$ and $\cos{\theta_e}$, respectively, are displayed in the middle panels. The shaded gray area indicates the kinematic region covered by the present calculations of response functions and densities. In the right panels, the solid (dashed) gray line indicates the electron longitudinal (transverse) kinematic factor appearing in Eq.~(\ref{eq:cross_section}). The dashed green vertical lines highlight the positions of the peaks. See text for details. }
\label{fig:q_values}
\end{figure*}

Response densities provide additional information on the two-nucleon final state immediately after the interaction with the external probe occurred. For example, we can analyze the STA density in terms of $pp$ and $nn$ pairs' contributions at two-body kinematics of experimental interest. The contributions from different particle identities have been explored in various electron scattering experiments~\cite{Hen:2014nza,Arrington:2011xs,Fomin:2017ydn} in the back-to-back kinematics, where the struck pairs have initial center of mass momentum equal to zero, implying $E_{\rm cm}\approx |{\bf P^\prime}_{\rm c.m.}|^{2} /(4\, m) = |{\bf q}|^2 /(4\, m)$. As an example, we analyze the response density at momentum transfer $|{\bf q}|= 570$ MeV (illustrated in Fig.~\ref{fig:pairs}), where the back-to-back configuration is realized at values of the center of mass energy $E_{\rm cm}\sim 87$ MeV. The longitudinal (upper panel) and transverse (lower panel) response densities at this fixed value of center of mass energy are displayed in Fig.~\ref{fig:pairs} as a function of the relative energy. Additionally, we show the contributions due to $pp$ (dashed red line), $nn$ (dashed-dotted blue line) pairs,
total (solid black line), one-body (solid green line) and interference (solid orange line) results. In the longitudinal case, the response is almost entirely due to one-body charge operators.
The $nn$ contribution is negligible with respect to the $pp$ one, since it is proportional to the neutron charge operator squared. The $pp$ term is enhanced by a factor of $\sim 1.3$ with respect to the $pp$ contribution found in $^4$He within the same kinematic configuration~\cite{Pastore:2019urn}. This result can be understood in terms of pair counting; {\it i.e.}, the number of $pp$ pairs over the total number of pairs is $1/6$ in $^4$He versus $15/66$ in $^{12}$C, which leads to the observed enhancement.
The transverse density is shown in the bottom panel of Fig.~\ref{fig:pairs}. Here, the $pp$ and $nn$ contributions are primarily due to one-body current operators and are proportional to proton and neutron magnetic form factors squared, respectively, which explains the observed enhancement of the $pp$ contribution versus the $nn$ one. Two-body contributions, primarily driven by $np$ pairs, become predominant as the relative energy increases.

\subsection{Response functions}

Longitudinal and traverse response functions, obtained from the densities discussed above, are shown in Fig.~\ref{fig:functions} for values of momenta $|\mathbf{q}|=300, \, 380$, and $570$ MeV. This choice of momenta allows us to compare with both the exact GFMC results of Ref.~\cite{Lovato:2016gkq} and the experimental data of Refs.~\cite{Barreau:1983ht,JOURDAN1996117}. Note that, while the response functions at $|\mathbf{q}|=300$ and $570$ MeV/$c$ are obtained from an integration of the calculated densities over the relative and center of mass energies, the response function at $|\mathbf{q}|=380$ MeV/$c$ is obtained implementing the interpolation scheme described in Sec.~\ref{sec:cross}. In the figure, we display the error bands that account for the statistical errors inherent to the stochastic calculations.
Overall, in the analyzed kinematics, we find a good agreement with both the GFMC results and the experimental data.
We highlight the impact of two-body currents in Fig.~\ref{fig:res_fun_2b}, where we show the longitudinal and transverse response functions at  $|\mathbf{q}|=570$ MeV/$c$. In particular,
we show the one-body contributions (dashed lines), and total responses (solid lines), which include the contributions of the one- and two-body interference and the pure two-body terms. In the figure, the longitudinal and transverse responses are indicated by blue and red lines, respectively. Two-body currents provide a $\sim20$ \% contribution to the total transverse response function.
The two-body corrections to the longitudinal response function are less than $3$ \%.

\subsection{Cross sections}
\label{sec:results_cross}

\begin{figure*}
\centering
    \begin{subfigure}[t]{\textwidth}
    \includegraphics[width=0.32\linewidth]{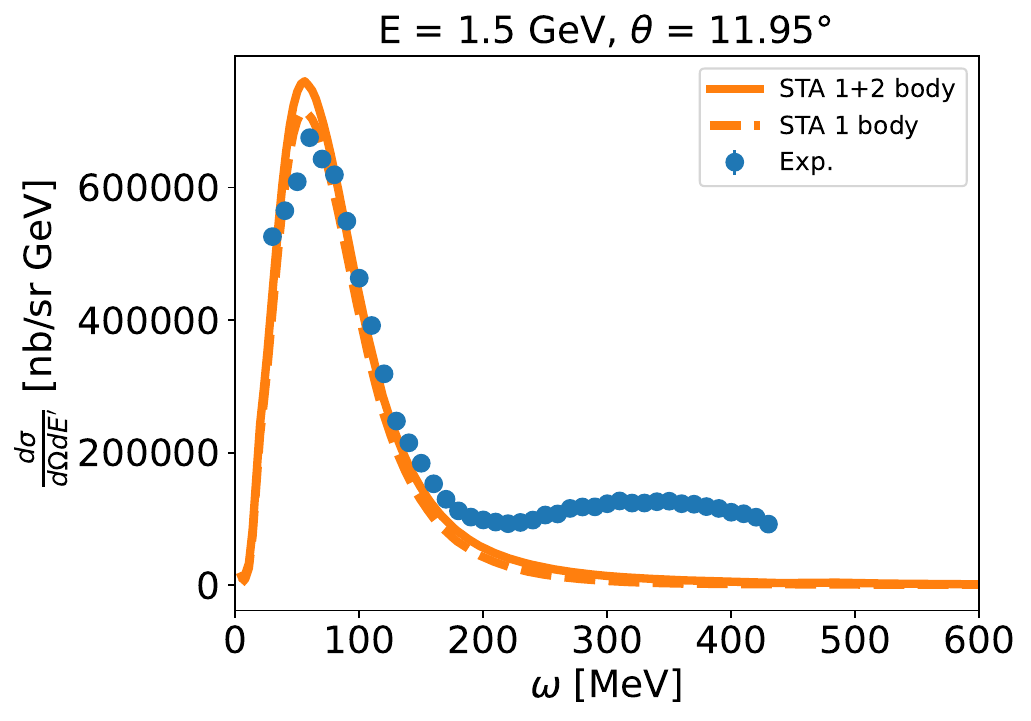} 
    \includegraphics[width=0.32\linewidth]{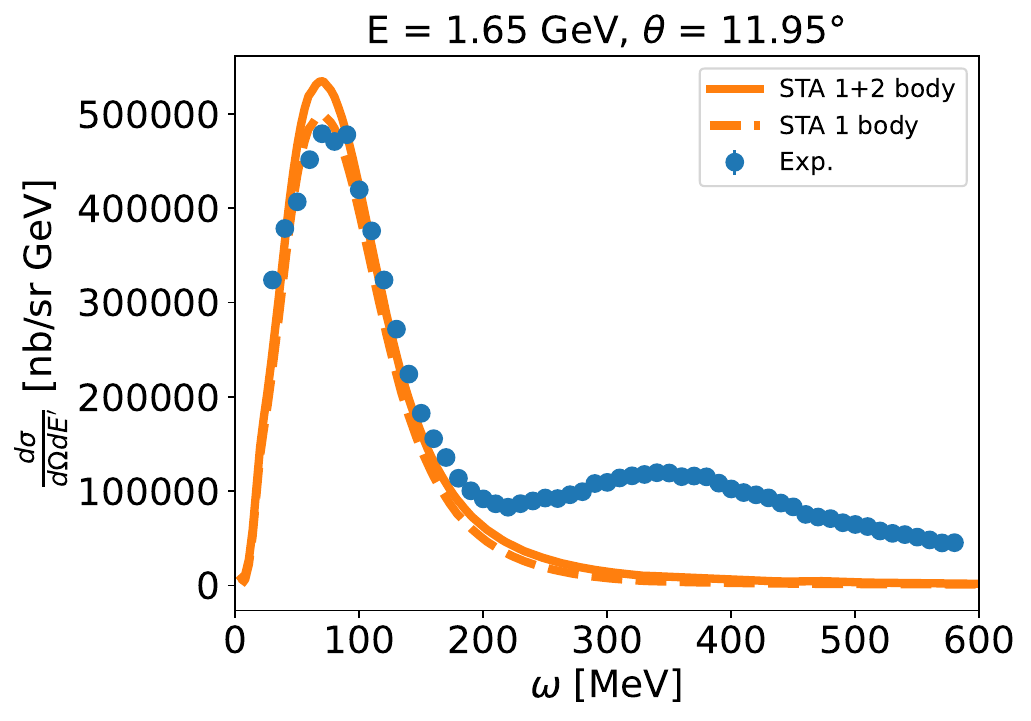}
    \includegraphics[width=0.32\linewidth]{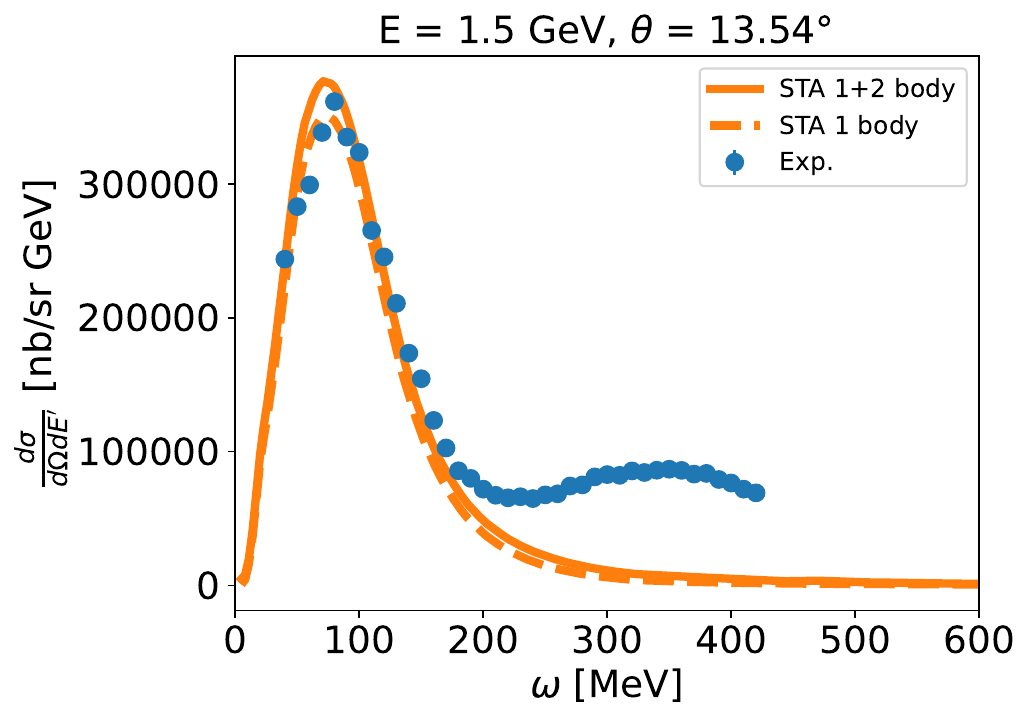}
    \end{subfigure}
    \begin{subfigure}[t]{\textwidth}
    \includegraphics[width=0.32\linewidth]{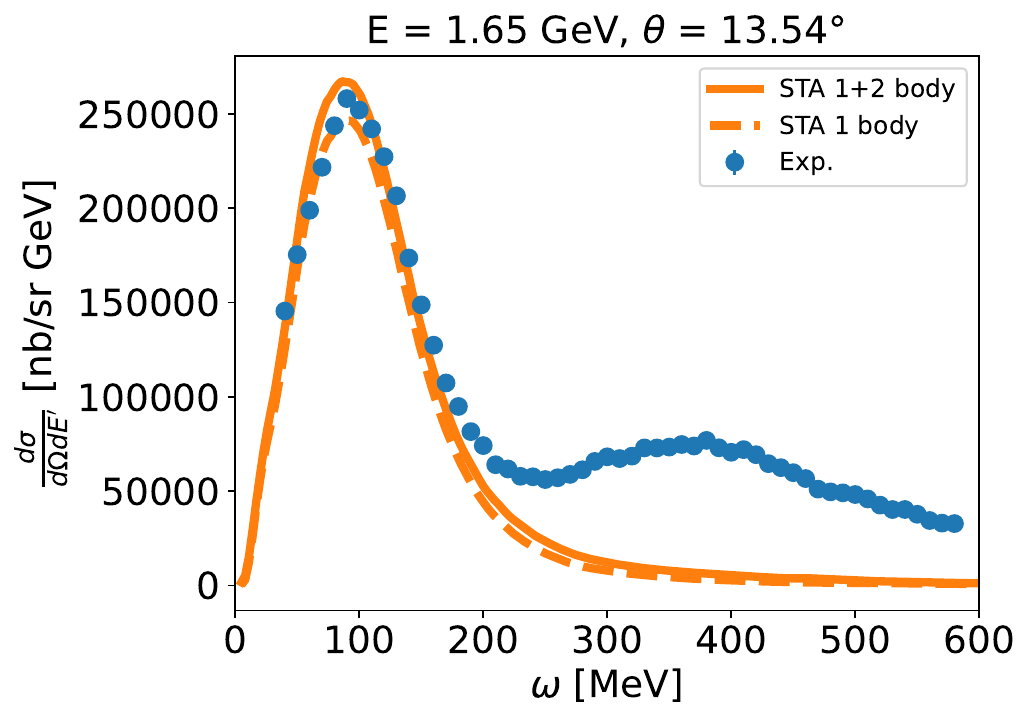}
    \includegraphics[width=0.32\linewidth]{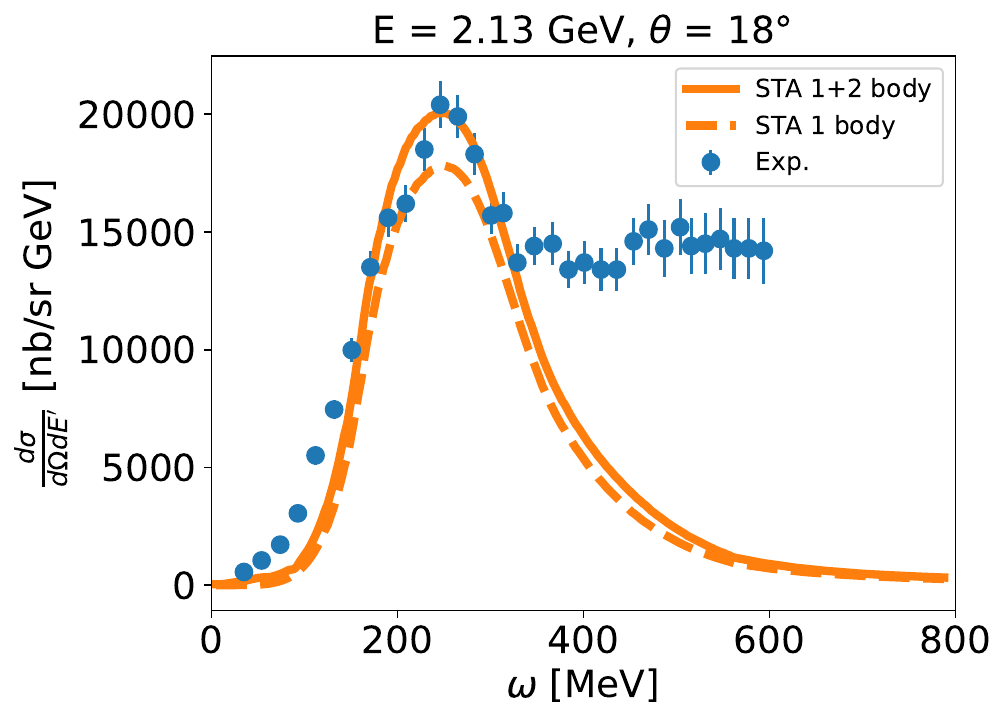}
    \includegraphics[width=0.32\linewidth]{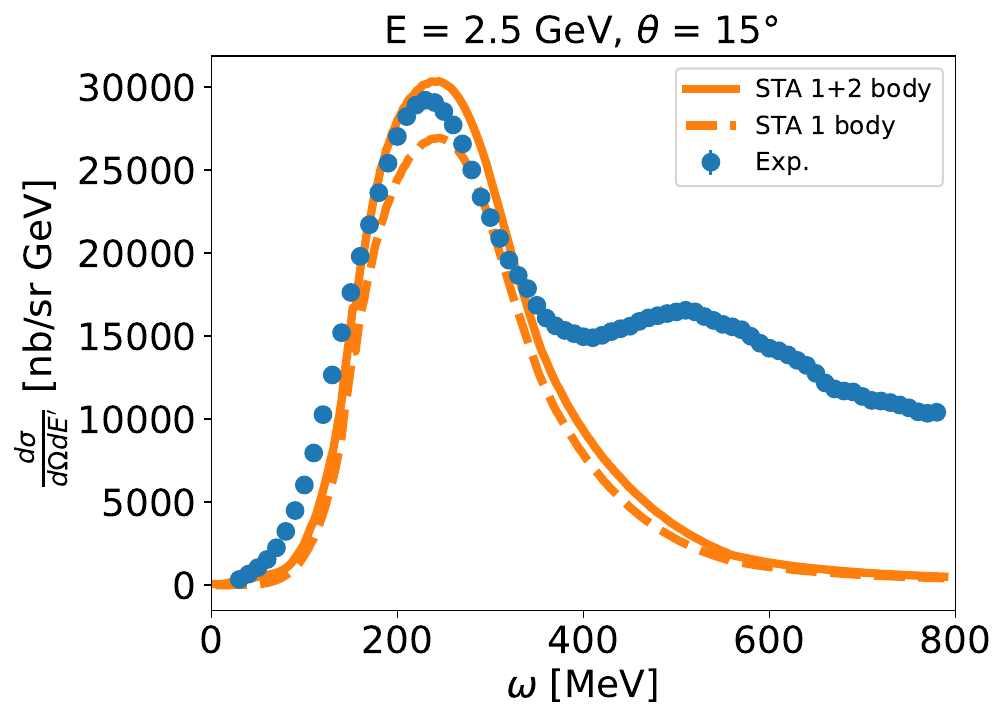} 
    \end{subfigure}
    \begin{subfigure}[t]{\textwidth}
    \includegraphics[width=0.32\linewidth]{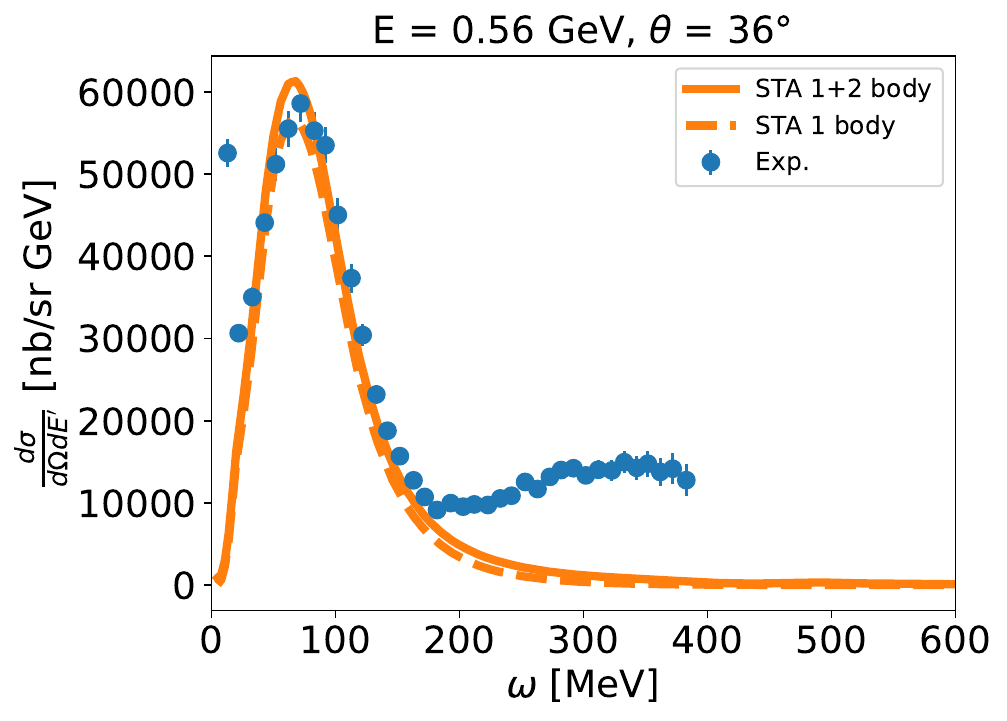} 
    \includegraphics[width=0.32\linewidth]{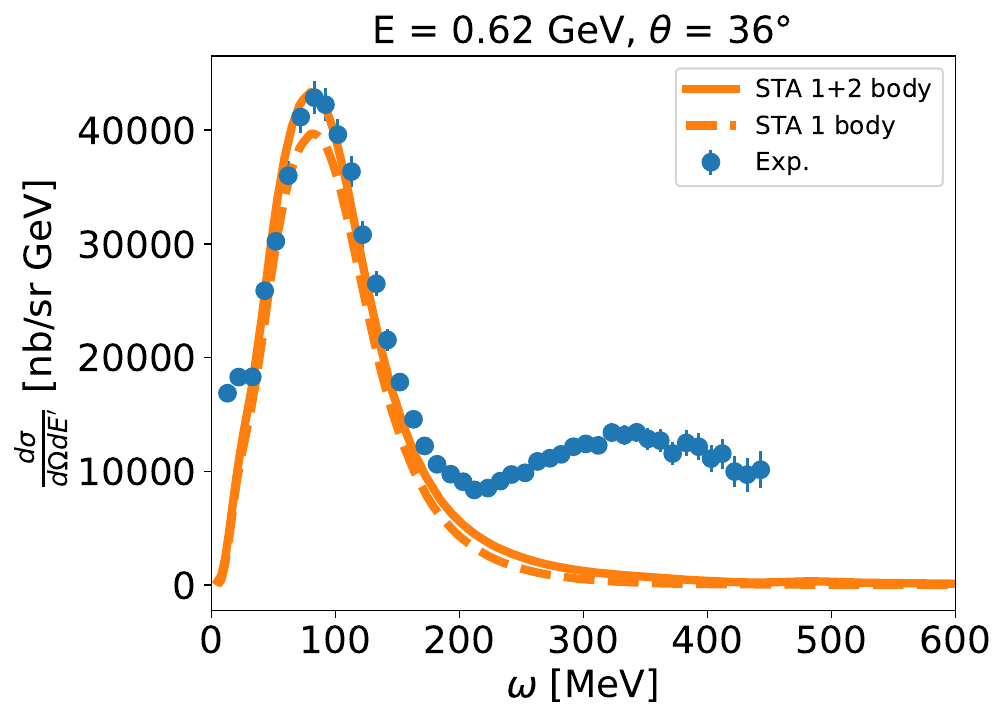}
    \includegraphics[width=0.32\linewidth]{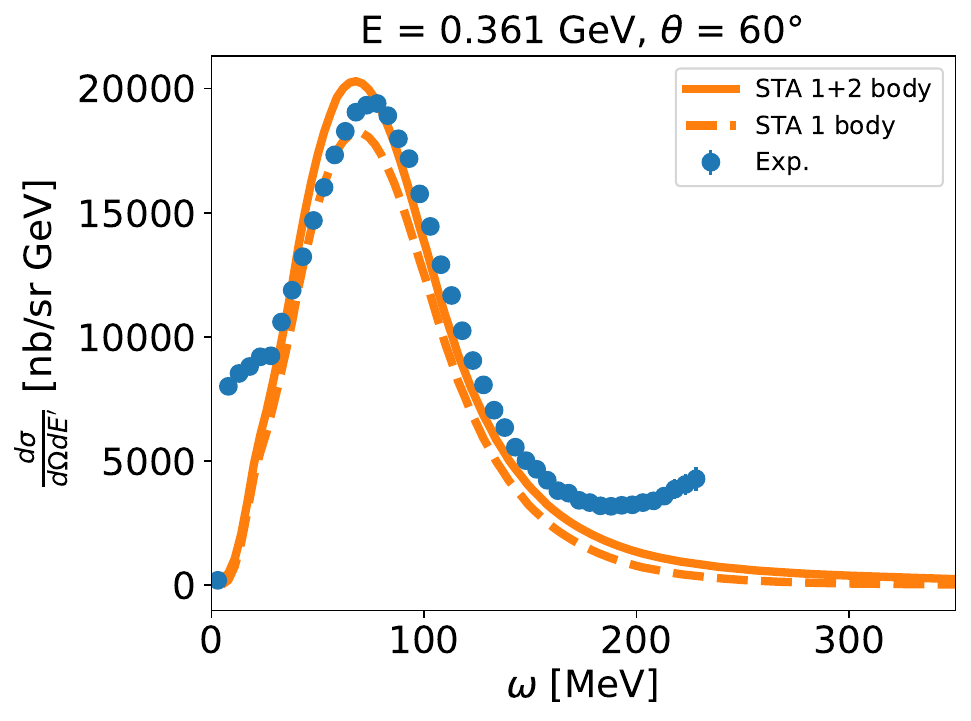} 
    \end{subfigure}
    \begin{subfigure}[t]{\textwidth}
    \includegraphics[width=0.32\linewidth]{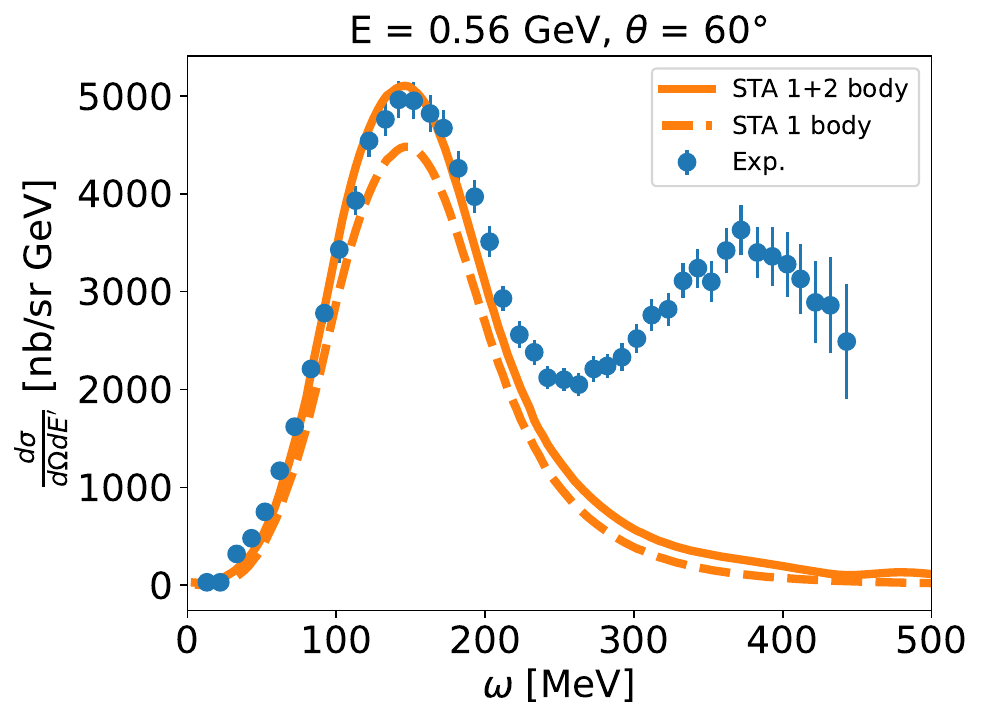} 
    \includegraphics[width=0.32\linewidth]{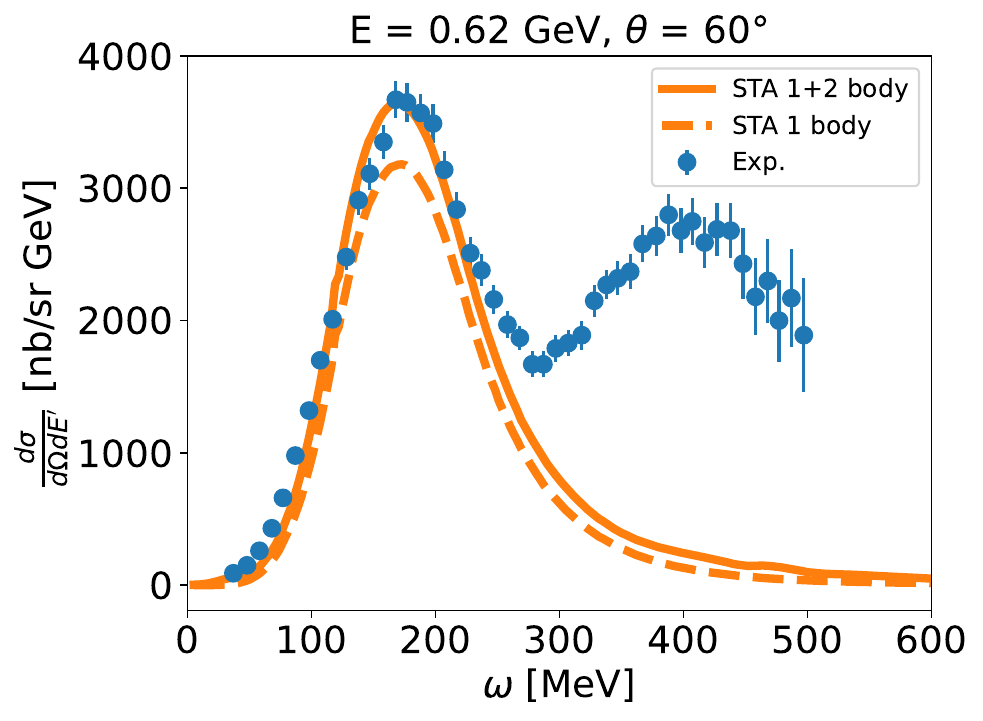} 
    \includegraphics[width=0.32\linewidth]{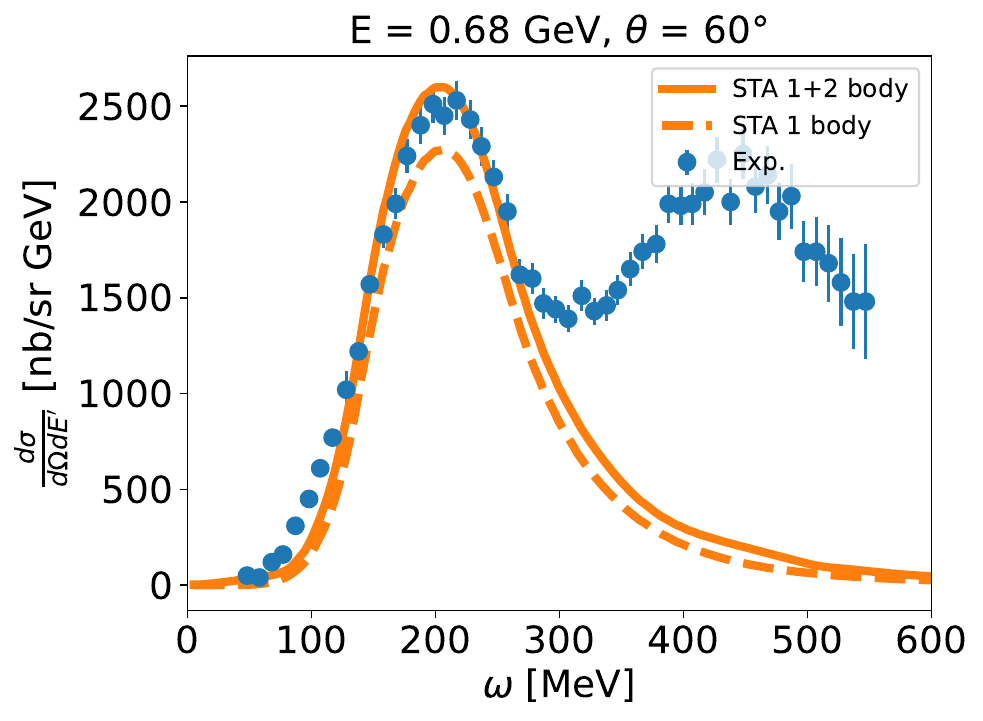}
    \end{subfigure}
\caption{Inclusive double-differential cross sections for electron scattering on \carbon, for various values of the incoming electron beam energy and scattering angle. Experimental data for $\theta=11.95\degree,13.54\degree$ is from~\cite{Baran:1988tw},
for $\theta=15\degree$ is from~\cite{Zeller:1973ge},
for $\theta=18\degree$ is from~\cite{Bagdasaryan:1988hp}, and
for $\theta=36\degree,60\degree$ is from~\cite{Barreau:1983ht}}
\label{fig:cross}
\end{figure*}

We use the interpolation scheme discussed in Sec.~\ref{sec:cross} to generate response functions on a $1$ MeV/$c$ spaced grid of transferred momenta in the range $[300,800]$ MeV/$c$. The cross sections obtained from them are compared with the experimental data from Refs.~\cite{Zeller:1973ge,Bagdasaryan:1988hp,Baran:1988tw,Barreau:1983ht}, collected in a tabular form in Refs.~\cite{Benhar:2008,Virginiaarchive}. We analyze experimental results with values of incoming electron energy and electron scattering angle in the ranges $E_e\sim[0.3,2.5]$ GeV and $\theta_e\sim[11^{\circ},60^{\circ}]$. In Fig.~\ref{fig:q_values}, we discuss in more detail three selected kinematics, namely $(E_e,\theta_e)=(0.62\, {\rm GeV}, 60^{\circ})$, $(1.5\, {\rm GeV}, 13.54^{\circ})$, $(1.108\, {\rm GeV}, 37.5^{\circ})$, and highlight some basic features displayed by our calculations. In particular, for each row, we show a comparison of the inclusive double differential cross section (solid orange line) with the experimental data (blue symbols) in the first panel; the second panel shows the values of momentum and energy transfer, $|{\bf q}|$ and $\omega$, corresponding to the given electron energy and scattering angle, $E_e$ and $\theta_e$. These are highlighted by the solid gray line. Additionally, the shaded region indicates the region covered by the calculated response functions. In the last panel, we report the values of the electron kinematic factors, $A_L$ and $A_T$, entering the cross section--see Eq.~(\ref{eq:cross_section}). The vertical dashed green line serves to guide the eyes and highlights the position of the peak. We find very good agreement between our calculations and the experimental data, in particular for kinematics that require the evaluation of response functions in the range $|\mathbf{q}|\sim [300,600]$ MeV$/c$. This is the case, for example, for the kinematics shown in panels~(\ref{fig:q_values1a})
and~(\ref{fig:q_values1b}) of Fig.~\ref{fig:q_values}, corresponding to $(E_e,\theta_e)=(0.62\, {\rm GeV}, 60^{\circ})$ and  $(1.5\, {\rm GeV}, 13.54^{\circ})$, respectively. The former requires response functions in the range $|\mathbf{q}|\sim [550,600]$ MeV$/c$ with $A_L\sim A_T$ at the peak, while the latter uses response functions in the range $|\mathbf{q}|\sim [350,450]$ MeV$/c$ with $A_L\sim 2\, A_T$ at the peak. 

Panel~(\ref{fig:q_values1c}) displays results at $(E_e,\theta_e)=(1.108\, {\rm GeV}, 37.5^{\circ})$. At these kinematic, response functions with $|\mathbf{q}|\gtrsim 600$ MeV$/c$ are required to calculate the cross sections.
In particular, values of $|\mathbf{q}|\sim [650,750]$ MeV$/c$ contribute to the main peak of the cross section, which is overpredicted by the theoretical calculation. This discrepancy is indicative of the lack of a proper relativistic treatment of the interaction of the external probe with the correlated pairs of nucleons at the vertex, a feature also observed in previous studies on $A=3$ systems~\cite{Andreoli:2022}.

We conclude this section by showing in Fig.~\ref{fig:cross} results obtained for incoming electron energies in the range $E_e\sim[0.3,2.5]$ GeV.
Each panel in Fig.~\ref{fig:cross} shows the 
contribution to the total cross section due to two-body physics, where the one-body term (includes both diagonal and off-diagonal components) is represented by the dashed orange line, while the total cross section comprehensive of the two-body correction (includes both the one- and two-body interference term along with the pure two-body component) is shown by the solid orange line.
At these kinematics, given the combination of longitudinal and transverse responses coming from equation~\ref{eq:cross_section}, two-body effects provide an enhancement at the peak between 5 \% and 15 \%, for the ranges of energies and angles considered.

Overall the STA accurately explains experimental data in the quasi-elastic peak region, while it fails at higher values of $\omega$ for which the inclusion of resonances and meson-production, currently not accounted for in the theory, is required.

\section{Conclusions}
\label{sec:conclusions}

In this work, we performed VMC calculations within the STA of inclusive electron scattering from $^{12}$C. Specifically, we calculated \carbon\, electromagnetic longitudinal and transverse response densities and functions for five values of momentum transfer in the range $|\mathbf{q}|=300-800$ MeV$/c$. In the calculation, we accounted for two-body physics at the vertex, namely, two-body correlations induced by the AV18 two-nucleon potential, and consistent two-nucleon electromagnetic currents. The VMC ground state is fully correlated via the AV18 two-nucleon and Urbana-X three-nucleon potentials. We found a very good agreement between the STA results, the experimental data, and the  GFMC calculations of response functions available at $|{\bf q}|=300$, $380$ and $570$ MeV$/c$.  

We developed an interpolation scheme that conserves the sum rules, and used it to generate response functions on a fine grid of transferred momenta from the calculated and sparse ones. This allowed us to calculate  electron-\carbon\, inclusive cross sections for electron energies in the range $E_e \sim [0.3,2.5]$ GeV. 
We found that the STA gives a very good description of the quasi-elastic peak, provided that the transfer momentum lies in the range $|{\bf q}|\sim [300,600]$ MeV$/c$. For values of $|{\bf q}|\gtrsim 600$ MeV$/c$, a proper inclusion of relativistic effects, entering both the kinematics at the vertex and the interaction of the external probe with the correlated pairs of nucleons, is required for an accurate description of the data.  The factorization scheme adopted in the STA for the final hadronic state will allow us to overcome these limitations, as the energies and momenta of the two nucleons participating in the scattering process are directly accessible. A work on the inclusion of relativistic effects, using relativistic expressions for the electromagnetic currents and of the kinematics of the two nucleons involved in the scattering is underway~\cite{Weiss:inpreparation}. Additionally, the STA formalism, explicitly considering two active nucleons in the final state, is also amenable to the inclusion of pion production channels. A similar implementation, {\it albeit} developed within the spectral function formalism, has been adopted, {\it e.g.}, in Ref.~\cite{Rocco:2019}.

These studies on electron-nucleus scattering are relevant to the study of neutrino-nucleus scattering. They allow us to test the nuclear model, and assess the relevance of many-nucleon effects, including correlations and currents, that play an important role also in scattering induced by neutrinos. The group is currently pursuing STA calculations of electroweak response densities. The latter provides important additional information on two-nucleons final states.  
Nuclear electromagnetic response functions obtained within the STA have been implemented into the GENIE Monte Carlo event generator~\cite{Barrow:2021}, through a \textit{hadron tensor} interface~\cite{Gardiner2019}. Future work will directly implement the multidimensional information contained in the nuclear response densities.

In this work, the STA has been implemented into the VMC computational scheme, currently limited to the study of $A=12$ systems. Calculations in heavier systems will be possible using next-generation high-performance computing systems. Additionally, the STA can be exported to the Auxiliary Field Diffusion Monte Carlo method~\cite{Schmidt:1999, Carlson:2014vla}  that can address large nuclear systems up to $A\sim 20$. Work along these lines is in progress.

\section*{Acknowledgments}
We thank A.~Lovato and N.~Rocco for interesting discussions at various stages of this work and for sharing with us the GFMC results.

 This work is supported by the US Department of Energy under Contracts No.~DE-SC0021027 (L.A., G.~B.~K. and S.~P.), DE-AC02-06CH11357 (R.B.W.), a 2021 Early Career Award number DE-SC0022002 (M.~P.), the FRIB Theory Alliance award DE-SC0013617 (M.~P.), the NUCLEI SciDAC program under Award number DE-SC0023495 (L.A., S.P., M.P. and R.B.W.), and the Neutrino Theory Network (S.P.). G.~B.~K. would like to acknowledge support from the U.S. DOE NNSA Stewardship Science Graduate Fellowship under Cooperative Agreement DE-NA0003960. L.A. would like to acknowledge support from the McDonnell Center for the Space Sciences, the Visiting Scholars Award Program of the Universities Research Association, and the U.S. Department of Energy Contract No.~DE-AC05-06OR23177, under which Jefferson Science Associates, LLC operates Jefferson Lab.
 We thank the Nuclear Theory for New Physics Topical Collaboration, supported by the U.S.~Department of Energy under contract DE-SC0023663, for fostering dynamic collaborations.

 The work of J.C. and S.G. was supported by the U.S.Department of Energy Office of Science, Scientific Discovery through Advanced Computing(SciDAC) NUCLEI program,
 and by  the US DOE/SC nuclear theory program through LANL. LANL is operated by Triad National Security, LLC, for NNSA of US DOE (Contract No. 89233218CNA000001. 
 
The many-body calculations were performed on the parallel computers of the Laboratory Computing Resource Center, Argonne National Laboratory, the computers of the Argonne Leadership Computing Facility (ALCF) via the INCITE grant ``Ab-initio nuclear structure and nuclear reactions'', the 2019/2020 ALCC grant ``Low Energy Neutrino-Nucleus interactions'' for the project NNInteractions, the 2020/2021 ALCC grant ``Chiral Nuclear Interactions from Nuclei to Nucleonic Matter'' for the project ChiralNuc, the 2021/2022 ALCC grant ``Quantum Monte Carlo Calculations of Nuclei up to $^{16}{\rm O}$ and Neutron Matter" for the project \mbox{QMCNuc}, and by the National Energy Research
Scientific Computing Center, a DOE Office of Science User Facility
supported by the Office of Science of the U.S. Department of Energy
under Contract No. DE-AC02-05CH11231 using NERSC award
NP-ERCAP0027147.

\bibliography{bibliography}
\end{document}